\documentclass[a4paper,11pt]{article}
\pdfoutput=0 

\usepackage[T1]{fontenc} 
\relax
\usepackage{tabularx}
\usepackage{color}
\usepackage{colordvi}
\usepackage[normalem]{ulem}
\usepackage{color}

\usepackage{graphicx}
\usepackage{epsfig}
\usepackage{graphicx}
\usepackage{longtable}

\usepackage{jheppub} 

\usepackage{hyperref}

\usepackage{amsmath,amssymb,bm}
\usepackage{latexsym}

\def\Comment#1{}
\newcommand{\bean}{\begin{eqnarray*}}
\newcommand{\eean}{\end{eqnarray*}}

\newcommand{\gapproxeq}{\lower
.7ex\hbox{$\;\stackrel{\textstyle >}{\sim}\;$}}
\newcommand{\lapproxeq}{\lower
.7ex\hbox{$\;\stackrel{\textstyle <}{\sim}\;$}}

\newcommand\lsim{\mathrel{\rlap{\lower4pt\hbox{\hskip1pt$\sim$}}
    \raise1pt\hbox{$<$}}}
\newcommand\gsim{\mathrel{\rlap{\lower4pt\hbox{\hskip1pt$\sim$}}
    \raise1pt\hbox{$>$}}}
\newcommand{\ba}{\begin{array}}
\newcommand{\ea}{\end{array}}
\newcommand{\nn}{\nonumber}

\newcommand{\be}{\begin{equation}}
\newcommand{\ee}{\end{equation}}
\newcommand{\bear}{\begin{eqnarray}}
\newcommand{\eear}{\end{eqnarray}}

\newcommand{\ket}{\,\rangle}
\newcommand{\bra}{\langle \,}
\newcommand{\cO}{{\cal O}}

\newcommand{\mL}{\mathcal{L}}

\newcommand{\mF}{\mathcal{F}}
\newcommand{\mG}{\mathcal{G}}
\newcommand{\mH}{\mathcal{H}}

\newcommand{\mM}{\mathcal{M}}

\newcommand{\mP}{\mathcal{P}}

\newcommand{\Frac}[2]{\frac{\displaystyle #1}{\displaystyle #2}}
\newcommand{\Int}{\displaystyle{\int}}
\def\bat{\begin{array}{cc}}

\newcommand{\2}{\lambda_2}

\vspace{0.5cm}

\title{\boldmath
Refining the scalar and tensor contributions in $\tau\to \pi\pi\pi\nu_\tau$ decays
}

\author[1]{Juan Jos\'e Sanz-Cillero,}

\affiliation[1]{Departamento de F\'\i sica Te\'orica I,
Universidad Complutense de Madrid, E-28040 Madrid, Spain }
\emailAdd{jjsanzcillero@ucm.es}

\author[2]{Olga Shekhovtsova}
\affiliation[2]{NSC KIPT Akhiezer Institute for theoretical Physics, 61108 Kharkov, Ukraine}
\emailAdd{shekhovtsova@kipt.kharkov.ua}

\abstract{
In this article we analyze the contribution from intermediate spin--0
and spin--2 resonances to the $\tau\to\nu \pi\pi\pi$ decay
by means of a chiral invariant Lagrangian incorporating these mesons. In particular, we
study the corresponding axial-vector form-factors.
The advantage of this procedure with respect to previous analyses is that
it incorporates chiral (and isospin) invariance and, hence,
the partial conservation of the axial-vector current. This ensures
the recovery of the right low-energy limit, described by chiral perturbation theory,
and the transversality of the current in the chiral limit at all energies.
Furthermore, the meson form-factors are further improved by
 requiring appropriate QCD high-energy conditions.
We end up with a brief discussion on its implementation
in the Tauola Monte Carlo and the prospects for future analyses
of Belle's data.

}

\begin{document}
\today \\
\maketitle
\flushbottom

\newpage
\section{Introduction}

The aim of this letter is to provide a coherent description of the impact of scalar ($J^{PC}=0^{++}$) and tensor ($J^{PC}=2^{++}$) mesons in tau decays with three pions in the final state.
The four targets of this theoretical analysis are
\begin{itemize}

\item{\bf Chiral invariance and (partial) axial-vector current conservation: }
the chiral invariant Lagrangian framework considered in this letter ensures the right QCD symmetries and leads to
a hadronic matrix element which is transverse ($\partial_\mu J_A^\mu =0$)
in the chiral limit $m_q\to 0$ and where longitudinal corrections come naturally suppressed by $m_q$.
In addition, as isospin is a subgroup of the chiral symmetry, our chiral invariant
Lagrangian approach yields the right relation between the $\pi^0\pi^0\pi^-$ and $\pi^-\pi^-\pi^+$
tau decay form-factors, prescribed by isospin symmetry~\cite{Girlanda:1999fu}, without any further
requirement. Likewise, we will be always assuming the other symmetries of QCD,
parity and charge conjugation.~\footnote{These assumptions also imply $G$-parity conservation, which is a
combination of charge conjugation and isospin symmetry.}

\item{\bf Low-energy limit:}
the construction of a general chiral invariant Lagrangian that includes the chiral pseudo-Goldstones and the
meson resonances ($1^{++}$ axial-vector, $2^{++}$ tensor, etc.)
ensures the right low-energy structure and the possibility to match the low-energy effective field theory (EFT)
of QCD, Chiral Perturbation Theory ($\chi$PT).

\item{\bf On-shell description}:
previous works, in spite of neglecting the previous principles,
have performed a fine work in describing the decays through axial-vector and tensor resonances
when their intermediate momenta are near their mass shell~\cite{Castro:2011zd,CLEO:1999}.
Our outcome reproduces these previous results
when the momentum $k$ flowing through the intermediate resonance propagator becomes on-shell, this is,
when $k^2\approx M_R^2$ (for the corresponding $k$ and $M_R$). The chiral invariant Lagrangian ensures
that the previous properties are fulfilled also off-shell ($k^2\neq M_R^2$).

\item{\bf High-energy limit:}
by imposing high-energy conditions and demanding the behaviour prescribed by QCD
for the form-factors at short-distances we will constrain the resonance parameters.
Implementing these QCD principles will make our theoretical determination
phenomenologically predictive.

\end{itemize}

This resonance chiral theory (R$\chi$T) approach to the $3\pi$ tau decay was considered in the past
taking into account the impact of the vector and axial-vector
resonances~\cite{Dumm:2009va}. The corresponding current has been implemented into the Monte Carlo event generator Tauola~\cite{Shekhovtsova:2012ra}.
The comparison with the unfolded distributions from the preliminary BaBar Collaboration analysis~\cite{Nugent:2013ij}
for the three-prong mode has demonstrated
the mismatch in the low-energy part of the two-pion spectrum~\cite{Shekhovtsova:2012ra}
and was associated with the lack of the scalar meson multiplet in the original R$\chi$T current~\cite{Dumm:2009va}.
The scalar resonance contribution was later added
to the three pion current phenomenologically in Ref.~\cite{Nugent:2013hxa}.
However, the corresponding part does not obey
isospin symmetry~\cite{Finkemeier:1996,Girlanda:1999fu}
and, as a result, does not reproduce the proper chiral low-energy
behaviour (see the discussion in Sec.~\ref{sec:general} and App.~\ref{app.ChPT}).

This letter focuses on the impact of the lowest scalar ($\sigma$ and $f_0(980)$) resonances and the isosinglet tensor
 $f_2(1270)$,   which may be directly produced from the $W^-$ or generated via an intermediate pion or
 an $a_1$ state. Also we discuss the implementation of the associated currents into Tauola
 and present
 an estimate of tensor and scalar contributions to the three-pion partial width.
In Sec.~\ref{sec:general}, one finds the general formulae for the three-pion
axial-vector form-factor (AFF): the Lorentz structure decomposition
and the isospin relation between $\pi^-\pi^-\pi^+$ and $\pi^0\pi^0\pi^-$ channels.
In order to avoid any possible double-counting we have separated the contributions to the three-pion AFF
in the following way:
1) previous $3\pi$-AFF computations~\cite{Dumm:2009va,Shekhovtsova:2012ra} incorporate
the diagrams including vector resonance exchanges and non-resonant
contributions from the $\cO(p^2)$ $\chi$PT Lagrangian~\cite{rcht};
2) Sec.~\ref{sec:S} provides the contribution to the $3\pi$-AFF from diagrams with scalar exchanges;
3) the contribution due to spin--2 resonance exchanges is discussed in Sec.~\ref{sec:T}.
Sec.~\ref{sec:Tauola} is dedicated to the implementation in the Monte Carlo generator Tauola and some basic numerical
results. We provide the conclusions in Sec.~\ref{sec:conclusions} and some technical details
have been relegated to the Appendices.

\section{Axial-vector form-factor into three pions: general formulae}
\label{sec:general}

The matrix element of the tau-decay into the three pions is determined
 in terms of the transverse form-factors $\mF_1$, $\mF_2$ and $\mF_3$
and a longitudinal one
$\mF_P$:
\bear
\bra 3\pi |\bar{d}\gamma^\mu\gamma_5 u|0\ket &=&  H^{3\pi}(q^2,s_1,s_2)^\mu
\nn\\
&=&  i \, P_T^{\mu\nu} (q) \bigg[
\mF_1(s_1,s_2,q^2)\,\, (p_1 - p_3)_\nu \,\,+\,\, \mF_2(s_1,s_2,q^2) \,\,  (p_2-p_3)_\mu
\nn\\
&& +\,\, \mF_3(s_1,s_2,q^2) \,\,  (p_1-p_2)_\mu   \bigg]\,\,
+ \,\, i\, q_\mu \,\, \mF_P(s_1,s_2,q^2) \, , \label{eq.hadr-curr-3pions}
\eear
with $q=p_1+p_2+p_3$, $s_1=(p_2+p_3)^2$, $s_2=(p_3+p_1)^2$ and $s_3=(p_1+p_2)^2$,
and $P_T(q)^{\mu\nu}= g^{\mu\nu} -q^\mu q^\nu /q^2$.
The three transverse form-factors are linearly dependent and we will leave
only $\mF_1$ and $\mF_2$ as our basis.
The longitudinal form-factor $\mF_P$ vanishes in the chiral limit and is suppressed
by $m_\pi^2/q^2$~\cite{Dumm:2009va}.
Our formulae for the hadronic form-factors will be calculated in the isospin limit.
We will take $m_\pi = (m_{\pi^0} + 2 m_{\pi^+})/3$
and, in general, apply the relation $q^2 = s_1 + s_2 + s_3 -3m_\pi^2$
to express the form-factors in terms of the three independent kinematic variables $q^2,s_1,s_2$.

Bose symmetry implies that
\bear
\mF_1(s_1,s_2,q^2)  &=&  \mF_2(s_2,s_1,q^2) \, ,
\nn\\
\mF_P(s_1,s_2,q^2) &=&\, \mF_P(s_2,s_1,q^2)  \, ,
\eear
and therefore there are only two independent form-factors, e.g., $\mF_1$ and $\mF_P$.

Isospin symmetry relates the matrix elements with $\pi^-\pi^-\pi^+$ and $\pi^0\pi^0\pi^-$
final states~\cite{Girlanda:1999fu}:~\footnote{Isospin violation effects were found to be very suppressed in this decay, of the order of $0.4\%$
and $10^{-3}\%$, respectively for the $\pi^-\pi^-\pi^+$ and $\pi^0\pi^0\pi^-$
channels~\cite{Mirkes:1997ea}.
}
\bear
H_\mu^{--+}(p_1,p_2,p_3) &=& H_\mu^{00-}(p_3,p_2,p_1) +H_\mu^{00-}(p_3,p_1,p_2) \, .
\label{eq.isospin-rel1}
\eear
Thus, the form-factors for $\pi^-\pi^-\pi^+$ and $\pi^0\pi^0\pi^-$ are related in the form
\bear
\hspace*{-0.85cm} \mF_1^{\pi^-\pi^-\pi^+}(s_1,s_2,q^2) &=&
\mF_1^{\pi^0\pi^0\pi^-}(s_1,s_3,q^2) - \mF_1^{\pi^0\pi^0\pi^-}(s_2,s_3,q^2)
- \mF_1^{\pi^0\pi^0\pi^-}(s_3,s_2,q^2) \, ,
\label{eq.isospin-rel2A}
\\
\hspace*{-0.85cm} \mF_P^{\pi^-\pi^-\pi^+}(s_1,s_2,q^2) &=&
\mF_P^{\pi^0\pi^0\pi^-}(s_1,s_3,q^2)
 +  \mF_P^{\pi^0\pi^0\pi^-}(s_2,s_3,q^2) \, .
\label{eq.isospin-rel2}
\eear
It is also possible to revert this expressions and to express the $\pi^0\pi^0\pi^-$
matrix element
in terms of the $\pi^-\pi^-\pi^+$ (App.~\ref{app.relations})
but for sake of simplicity, from now on, we will always refer to the $\pi^0 \pi^0\pi^-$ form-factors
 and assume Eqs.~(\ref{eq.isospin-rel2A}) and (\ref{eq.isospin-rel2})
whenever the $\pi^- \pi^-\pi^+$
one is needed. The advantage of our chiral Lagrangian approach is that it implements by default
this isospin relation (and Bose symmetry, of course), as isospin is a subgroup
of the chiral group.

It is worth to stress that the $\pi^-\pi^-\pi^+$ and $\pi^0\pi^0\pi^-$  hadronic currents are
in general not the same~\cite{Finkemeier:1996,Pais:1960zz,CLEO-isospin}.
The diagrams with intermediate vector and axial-vector resonances give the same
$\mF_1(s_1,s_2,q^2)$
form-factor up to a global sign difference~\cite{Dumm:2009va}.
However, on the contrary to the approach therein,
tensor and scalar resonances generate contributions
to the $\pi^-\pi^-\pi^+$ and $\pi^0\pi^0\pi^-$ hadronic currents
with a different kinematical structure (determined by
Eqs.~(\ref{eq.isospin-rel2A}) and (\ref{eq.isospin-rel2})).
For further details on the isospin relation between channels see
Refs.~\cite{Girlanda:1999fu,Finkemeier:1996,Pais:1960zz} and App.~\ref{app.relations}.
In the next Sections we will focus on the three-pion tree-level production via
intermediate scalar and tensor resonances, which will be dressed with appropriate widths
when compared to data. Apart from this, we will not incorporate other one-loop contributions
like, e.g, the non-resonant triangular topologies with three internal propagators
(with the mesons $KKK^*$, $\pi\pi\rho$, etc.) and the external pions and $W$ connected
at the vertices.

\section{The decay $\tau \to \pi\pi\pi \nu_\tau$ through scalar resonances}
\label{sec:S}

We first consider the three-pion production via an intermediate state with a scalar $S$
and a pion.
If isospin and C-parity are conserved then G-parity
requires that the scalar resonance has isospin fulfilling $(-1)^I=+1$
--i.e., even isospin--, which in our case implies $I=0$.

The hadronic matrix element for the transition from an axial-vector current
into an isosinglet scalar $S$ and a pion has the general Lorentz structure~\cite{rcht-FFs}
\bear
\bra S_{I=0}(k) \pi^-(p)|\bar{d}\gamma^\alpha \gamma_5 u|0\ket &=&
- 2 i P_T(q)^{\alpha\nu}\,  p_\nu \, \mF^a_{S\pi}(q^2;k^2)
 \,\, +\,\, i\, q^\alpha \, \mH^a_{S\pi}(q^2;k^2)\, ,
\eear
where $q = k+ p$ and the scalar function $\mF^a_{S\pi}(q^2)$
provides AFF into $S\pi$ in the chiral limit, as $\mH^a_{S\pi}$ is suppressed by $m_\pi^2$
due to the partial conservation of the axial-vector current.
Here the
 isosinglet scalar $S_{I=0}$
refers to the resonance without $s\bar{s}$ component,
$S_{I=0}\sim u\bar{u}+d\bar{d}$, which we will relate with
the lightest  scalar isoscalar resonance, the $f_0(500)$ or $\sigma$.
We leave the discussion of the properness of this approach for a next Section:
here we will just assume  the large-$N_C$
framework~\cite{tHooft:1973alw,tHooft:1974pnl,Witten:1979kh}
and the phenomenological implementation will be later worked out.

\begin{figure}[!t]
\begin{center}
\includegraphics[width = 0.7\textwidth]{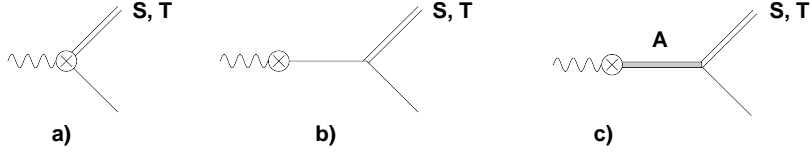}
\end{center}
\caption{{\small
Relevant diagrams for the hadronic tau decays into an isosinglet scalar $S$ and a pion
and its corresponding AFF
(similar to those for the decay into a isosinglet tensor $T$ and a pion).
Single straight lines stand for pions and the wavy line for the external axial-vector source
(from an incoming $W^-$).  }
}
\label{fig.diagr}
\end{figure}

In Fig.~\ref{fig.diagr}, we show the three relevant diagrams that must be taken into account
in the $S\pi$ production at large $N_C$ (and analogously later
in the production of a tensor resonance $T$ and a pion):
a) the direct production $W^- \to S \pi^-$;
b) the intermediate $\pi^-$ production $W^- \to \pi^- \to S  \pi^-$;
c) and the scalar production through an intermediate axial-vector resonance,
$W^- \to a_1 \to S \pi^-$.

\subsection{The R$\chi$T Lagrangian for scalar
fields}

The resonance Lagrangian has the generic structure
\bear
\mL_{\rm R\chi T} &=& \mL_{\rm non-R} \, +\, \sum_R \mL_R \,+\,
\sum_{R,R'} \mL_{R\, R'}\, +\, ...
\eear
which respectively contains operators without resonances, operators with one resonance field,
terms with two resonance fields, etc.
In the case of the tau decay into three pions through an intermediate scalar production,
the relevant chiral invariant Lagrangian consists of three parts:
\begin{itemize}
\item Operators with one resonance field~\cite{rcht}:
\bear
\mL_A &=& \Frac{F_A}{2\sqrt{2}} \bra A_{\mu\nu} f_-^{\mu\nu} \ket\, ,
\nn\\
\mL_S &=& c_d\bra S u_\mu u^\mu\ket + c_m\bra S\chi_+\ket\, ,
\eear
\item Operators with an axial-vector and a scalar field (which provides the $AS\pi$ vertex in
diagram c) in Fig.~\ref{fig.diagr})~\cite{rcht-FFs}:
\bear
\mL_{AS} &=& \lambda_1^{AS} \bra \{\nabla_\mu S , A^{\mu\nu}\} u_\nu\ket \, .
\label{eq.scalar-lagr}
\eear
 Operators of the $\mL_{AS}$ Lagrangian that do not contribute
to the $AS\pi$ vertex are not shown here~\cite{rcht-FFs}.
\item
Operators without resonance fields~\cite{Gasser:1983yg,Gasser:1984gg,rcht}:
\bear
 \mL^{(2)}_{\rm non-R}   &=& \Frac{F^2}{4}\bra u_\mu u^\mu +\chi_+\ket\, ,
\label{eq.non-R-lagr}
\eear
This non-resonant $O(p^2)$ Lagrangian generates the $W^-\to \pi^-$ transition vertex
in Fig.~\ref{fig.diagr}.b.
It also provides an $O(p^2)$  contribution without intermediate resonances to the $\pi\pi\pi$ AFF
which was accounted in previous analyses~\cite{Dumm:2009va}. Thus,
in order to avoid double counting,
we will not consider these non-resonant $\pi\pi\pi$ AFF diagrams.

\end{itemize}

For the axial-vector field $A_{\mu\nu}=A_{\mu\nu}^a \lambda^a/\sqrt{2}$
we have used the antisymmetric tensor representation~\cite{rcht,op6rxt}, with
\bear
A_{\mu\nu} &=&\left(\begin{array}{ccc}
0& a_1^+ & 0\\ a_1^- &0&0 \\ 0&0&0
\end{array} \right)_{\mu\nu} \,\,\,+\,\,\, ...
\eear
with the dots standing for the other axial-vector resonances of the multiplet,
which will not be relevant in the present study.
For the chiral tensors containing the
light pseudoscalars, the masses and the external vector and axial-vector source fields we used~\cite{chpt,rcht}
\bear
& U = u^2 =\exp\{ \pi^a \lambda^a/ F\}\, , \qquad
D_\mu U = \partial_\mu U - i r_\mu U + i U \ell_\mu\, , \qquad
u_\mu = i u^\dagger (D_\mu U) u^\dagger\, ,  &
\nn\\
& \chi_{\pm}= u^\dagger \chi u^\dagger \pm u \chi^\dagger u\, , \qquad
f_{\pm}^{\mu\nu} = u F_L^{\mu\nu} u^\dagger \pm u^\dagger F_R^{\mu\nu} u\, , \qquad
\nabla_\mu \cdot = \partial_\mu \cdot + [\Gamma_\mu , \cdot ]\, , &
\nn\\
& \Gamma_\mu =\Frac{1}{2} \left\{ u^\dagger (\partial_\mu - i r_\mu ) u + u(\partial_\mu - i \ell_\mu) u^\dagger
\right\} \, ,&
\eear
with the scalar-pseudoscalar source $\chi=2B_0$diag$(m_u,m_d,m_s)+...$
(the dots stand for terms not relevant for this calculation)
and $F_L^{\mu\nu}$ and $F_R^{\mu\nu}$ the field strength tensors
of the left and right sources,
respectively $\ell_\alpha$ and $r_\alpha$.
If we are only interested in the $W^\pm$ currents one takes
$\ell_\alpha = \frac{g}{\sqrt{2}} (W_\alpha^+ T_+ + {\rm h.c.})$ and $r_\alpha =0$,
with $T_+ = V_{ud} (\lambda^1+i\lambda^2)/2 + V_{us} (\lambda^4+i\lambda^5)/2$.
The $\pi^a$    generically  refer to the $SU(3)$ chiral pseudo-Goldstones ($a=1...8$).
At large $N_C$ (and for the non-strange current) this process only occurs for
the isosinglet scalar  $S_{I=0}\sim u \bar{u} +d\bar{d}$,
with no $s\bar{s}$ strange quark component:
\bear
S&=& \left(\begin{array}{ccc}
\Frac{S_{I=0}}{\sqrt{2}} & 0& 0 \\ 0 & \Frac{S_{I=0}}{\sqrt{2}} & 0 \\ 0&0&0 \end{array}\right)\,\,\, +\,\,\, ...
\eear
where the dots stand for other resonances in the multiplet not relevant for the present work.

\subsection{AFF into $S\pi^-$}

Our chiral invariant Lagrangian leads to the AFF prediction,~\footnote{
 There was a typo in the sign of the $F_A\lambda_1^{SA}$ term
 of $\mF^{a}_{S\pi}$ in Table A.2, App.~A in Ref.~\cite{rcht-FFs}.
It has been corrected in Eq.~(\ref{eq.Spi-AFF}).
The same applies to the later high-energy constraint~(\ref{eq.Spi+HE})
(the final constrained form-factor~(\ref{eq.Spi-AFF+HE})
remains nevertheless the same as in Ref.~\cite{rcht-FFs}).   }
\bear
\mF^a_{S\pi}(q^2  ;k^2   )&=& \Frac{2c_d}{ F_\pi  }
\quad + \quad
\Frac{\sqrt{2} F_A \lambda_1^{AS}}{  F_\pi  }\Frac{q^2}{M_A^2-q^2}
\, ,
\label{eq.Spi-AFF}
\\
\mH^a_{S\pi}(q^2  ;k^2   )&=&
\Frac{4}{F_\pi} \, \Frac{ m_\pi^2}{q^2 (q^2-m_\pi^2)}
\, \left[c_d (qp) + c_m q^2 \right] \, ,
\label{eq.Spi-AFF-2}
\eear
with $(q p)= (q^2+m_\pi^2-k^2)/2$, being $k^2=M_S^2$ for an on-shell scalar
(later, when this scalar is considered off-shell and decaying
in two pions with momenta $p_i$ and $p_j$ it will take
the value $k^2=(p_i+p_j)^2$).
The $c_m$ operator contributes through the $s$-channel pion exchange
to the longitudinal form-factor in Eq.~(\ref{eq.Spi-AFF-2}).~\footnote{
There is an indirect large-$N_C$
contribution to these form-factors
through the pion-wave function renormalization proportional to $m_\pi^2$
induced by the scalar 
Lagrangian~\cite{SanzCillero:2004sk}.
This effectively amounts to a replacement of $F$ by $F_\pi$,
as shown in~(\ref{eq.Spi-AFF}) and (\ref{eq.Spi-AFF-2}).
A similar thing happens in the other form-factors studied in the next Sections, where this pion-wave
function renormalization due to the scalars~\cite{SanzCillero:2004sk} is taken into account
in a similar way.}

\subsection{$3\pi$-AFF through an intermediate scalar resonance}
\label{sec:3pi-from-S}

Considering not only the $S\pi$ production but also the
subsequent decay   $S\to \pi \pi$  one obtains the corresponding contribution to the $\pi\pi\pi$-AFF.

Using the Lagrangian in Eqs.~(\ref{eq.scalar-lagr})--(\ref{eq.non-R-lagr}),
 we obtain the contribution from scalar resonance exchanges
to the $\pi^0\pi^0\pi^-$ AFFs defined in~(\ref{eq.hadr-curr-3pions}),
\bear\label{eq.ff-scal}
\mF_1^{\pi^0\pi^0\pi^-}(s_1,s_2,q^2)\bigg|_S &=&  \Frac{2}{3} \mF_{S\pi}^a(q^2 ;s_3  )\, \mG_{S \pi\pi}(s_3)\, ,
\\
\mF_P^{\pi^0\pi^0\pi^-}(s_1,s_2,q^2)\bigg|_S &=&   \mH^a_{S\pi}(q^2   ;s_3  ) \, \mG_{S\pi\pi}(s_3)\, ,
\eear
with $qp_j = (m_\pi^2 +q^2 -s_j)/2$. The $AS\pi$ form-factor is the previous one
in Eq.~(\ref{eq.Spi-AFF})
whereas propagation of the isosinglet $S$ and its decay into $\pi\pi$ gives
\bear
\mG_{S\pi\pi}(s_3) &=&
\Frac{\sqrt{2}}{ F_\pi^2   } \,\Frac{1 }{M_S^2 - s_3 }
\,  [ c_d(s_3-2m_\pi^2) + 2 c_m m_\pi^2]    \, .
\eear
Notice that we are giving the full result, including pion mass corrections
produced by our Lagrangian in Eqs.~(\ref{eq.scalar-lagr})--(\ref{eq.non-R-lagr}).
\footnote{
The function $\mG_{S\pi\pi}(s_3)$ is not the scalar form-factor
and, therefore, does not need to obey
asymptotic high-energy behaviour prescribed by QCD~\cite{Brodsky:1973kr}.
Notice that only on-shell hadron matrix elements are well-defined
and the off-shell behaviour is ambiguous as it can be modified through
field redefinitions in the hadronic generating functional~\cite{Gasser:1983yg,Gasser:1984gg}.
$\mG_{S\pi\pi}(s_3)$ just provides a) the on-shell decay $S\to \pi\pi$ (through its residue at $s_3=M_S^2$)
and b) the contribution to the $\pi\pi\pi$ AFF from topologies with an intermediate scalar
--either on-shell or off-shell--.
}

 Requiring that the contribution to the transverse component of
 the $\Pi_{AA}^{\mu\nu}(q)$ spectral function vanishes
implies that $\mF^a_{S\pi}(q^2) \longrightarrow 0 $ for $q^2\to\infty$ (see App.~\ref{app.optical-theorem}),
giving the constraint~\cite{rcht-FFs}
\bear
F_A\lambda_1^{AS} &=&    \sqrt{2} c_d\, ,
\label{eq.Spi+HE}
\eear
and the form-factor prediction
\bear
\mF^a_{S\pi}(q^2    ;s_3 ) &=& \Frac{2c_d}{ F_\pi   } \Frac{M_A^2}{M_A^2-q^2}
\, .
\label{eq.Spi-AFF+HE}
\eear
This high-energy constraint is similar to the asymptotic form-factor high-energy behaviour
prescribed by Brodsky-Lepage quark-counting rules~\cite{Brodsky:1973kr},
which imply, for instance, that the pion vector form-factor vanishes like $\sim 1/q^2$ at infinite momentum transfer~\cite{rcht,Brodsky:1973kr}.

The subsequent decay of the scalar into $\pi\pi$ is given by $\mG_{S\pi\pi}(s_3)$ and would provide
the absorptive $\pi\pi\pi$ contribution to Im$\Pi_{AA}^{\mu\nu}$. However,
in the narrow-width limit for $S$, the three-pion phase-space integral yields a delta function $\delta(s_3-M_S^2)$
that sets the $s_3$ value to $M_S^2$. Thus, the integral is factorized into the two-body integration of $|\mF^a_{S\pi}(q^2)|^2$
over the $S\pi^-$ phase-space and a constant angular integration over the phase-space of the two pions produced by the
scalar. Therefore, in this limit, the large $q^2$ behaviour of this three-pion contribution
to the spectral function is ruled by the form-factor $\mF^a_{S\pi}(q^2)$ in the way dictated
by Eq.~(\ref{eq.Spi-spectral-function}) (up to a global constant factor).
We will use this theoretical large--$N_C$ information and use it to constrain our form-factor even if we will later model it
in order to include important subleading effects in $1/N_C$ such as the $\sigma$ width.
\footnote{  Phenomenologically, in order to study the $a_1$ meson finite size effects,
Ref.~\cite{CLEO:1999}
considered an additional {\it ad hoc} exponential suppression factor
$\exp\{ - R^2  |\vec{p}_{\pi^-}|^2 /2\}$
in addition to the analogous $\mG_{S\pi\pi}(s_3)$ functions.
However, the fit to the experimental data did not show an essential difference between a zero
and non-zero value of $R$. As a result
 of this, the nominal fit shown therein was the one with $R = 0$
(for details see Section VI of \cite{CLEO:1999}).
Moreover,
these exponential factors do not have the right analytical structure
in the whole complex plane  and add  an exponentially divergent behaviour
for some complex directions at $|q^2|\to \infty$.
Likewise, this functional dependency may not come from a perturbative Lagrangian computation like the one
worked out in this article and will not be incorporated to our diagrammatic results.   }

The $S\pi$ AFF is then ruled by the $c_d$ coupling in the limit $m_\pi^2 \ll q^2$.
Even though its precise experimental value is still unclear,
most analyses agree on a value $c_d\sim 30$~MeV
(see~\cite{Escribano:2010wt} and references therein).
For a discussion on its numerical impact on the spectral distributions,
see Sec.~\ref{sec:Tauola}.

\subsection{Scalar resonance widths}
\label{sec:scalar-width}

The lightest isoscalar particle is the broad scalar $\sigma$, with
$M_\sigma^{\rm pole}= 441^{+16}_{-\, 8}$~MeV,
$\Gamma_\sigma^{\rm pole}=544^{+18}_{-25}$~MeV~\cite{CCL-sigma}.
It is thought to contain mostly just $u$ and $d$ quark components,  where the two--pion channel is its only kinematically allowed decay.
On the other hand,
 as it follows from its predominant decay into $K\bar{K}$,
the next scalar isosinglet, the $f_0(980)$,
is considered to have a large strange quark component,
being its $n\pi$ decay modes are suppressed.
However, for sake of completeness we will include both isoscalars into consideration.

A first approach to the physical QCD case  is provided  by the inclusion of
a $\sigma$--$f_0(980)$ splitting
through the substitution~\cite{Escribano:2006mb,Escribano:2010wt},
\begin{equation}
\Frac{1}{M_S^2\,-\,s}  \qquad  \longrightarrow \qquad
\Frac{\cos^2\phi_S}{
M_\sigma^2\,-\,s
 } \, \,+\,\, \Frac{\sin^2\phi_S}{M_{f_0}^2\,-\,s} \, ,
 \label{eq.MS-splitting}
\end{equation}
where $\phi_S$ is the scalar mixing angle.
For the $\sigma-f_0$ mixing we will use the numerical value
$\phi_S=-8^\circ$~\cite{Escribano:2006mb}.

Due to the $\sin^2\phi_S$ suppression the $f_0(980)$ produces a clearly
subdominant effect with respect to the impact of the broad~$\sigma$. However, the  comparison of the
modified R$\chi$T spectra~\cite{Nugent:2013hxa}~\footnote{
By \textit{modified} we mean a phenomenological approach
proposed in Sec.~II of~\cite{Nugent:2013hxa} to include the $\sigma$-meson
in the hadronic form-factors. }
with the unfolded distributions~\cite{Nugent:2013ij} from the preliminary BaBar Collaboration $\tau\to \nu_\tau \pi\pi\pi$~analysis
has shown a statistically significant mismatch:
the $\pi^+\pi^-$ experimental spectral function is well reproduced up to 1~GeV except for a small sharp bump concentrated at 980~MeV
which differs from the $f_0$-absent theoretical R$\chi$T expression by a few percent.
The inclusion of the $f_0$ and its occurrence here via the $\sigma-f_0$
mixing in Eq.~(\ref{eq.MS-splitting}) is expected to improve  the phenomenological description of the data.

\subsubsection{Incorporating the $\sigma$ meson width}
\label{sec:sigma-width}

So far in previous Sections we have carried on a large-$N_C$ computation
where one had an intermediate exchange of narrow-width scalars.
This approximation seems to be suitable for the $f_0(980)$. However,
the $\sigma$ meson is a broad resonance and the effect of its width
is non-negligible.
It is not our intention to enter here in the discussion of the $\sigma$ nature
but, rather,
to propose an improved parametrization of its effect
on the $\tau\to\nu\pi\pi\pi$ decay that incorporates the features described in the introduction.
For this, we follow the
successful analysis of subleading $1/N_C$ effects in scalar exchanges
in the $\eta'\to \eta \pi\pi$ process~\cite{Escribano:2010wt}:
after considering the scalar splitting
in~(\ref{eq.MS-splitting}),
we incorporate the ``dressed'' $\sigma$ propagator in a similar way
by performing the substitution
\begin{equation}
\label{eq.S-propagator2}
\Frac{1}{M_\sigma^2\,-\,s}  \qquad\longrightarrow\qquad     \Frac{1}{ M_\sigma^2\,-\,s\, -\, f_{\sigma}(s)\, -\, i M_\sigma \Gamma_\sigma(s)  }\, ,
\end{equation}
with
\bear
f(s)&=&
c_\sigma s^k\,{\rm Re} \overline{B}_0(s,m_\pi^2,m_\pi^2) \,=\,
\Frac{c_\sigma \, s^k}{16\pi^2}\left[2-\rho_\pi(s)\ln{\frac{\rho_\pi(s)+1}{1-\rho_\pi(s)}} \right]\, ,
\nn\\
M_\sigma \Gamma_\sigma(s)
&=&
c_\sigma s^k \,  {\rm Im} \overline{B}_0(s,m_\pi^2,m_\pi^2)
\,=\,
\Frac{c_\sigma \,\rho_P(s)\,  s^k }{16\pi}\, ,
\label{eq.S-self-energy}
\eear
in the fashion of Gounaris and Sakurai~\cite{GS-rho}
and the Chew and Mandelstam dispersive integral~\cite{Chew:1960iv}.
We will use the parameters
$M_\sigma$ and $c_\sigma$ tuned such that one recovers the
right position for the $\sigma$ pole,
$M_\sigma^{\rm pole}= 441^{+16}_{-\, 8}$~MeV,
${  \Gamma_\sigma^{\rm pole}=544^{+18}_{-25}   }$~MeV~\cite{CCL-sigma}.
The function,
\bear
\overline{B}_0(s,m_P^2,m_P^2)
&=&\frac{1}{16\pi^2}\left[2-\rho_P(s)\ln{\frac{\rho_P(s)+1}{\rho_P(s)-1}}  \right]
\nn\\
&&= \frac{1}{16\pi^2}\left[2-\rho_P(s)\ln{\frac{\rho_P(s)+1}{1-\rho_P(s)}}
\, +\, i \pi \rho_P(s) \right]
\, ,
\eear
is  \ \ \ the  \ \ \ subtracted  \ \ \ two--point \ \ \ Feynman  \ \ \ integral  \ \ \ ($\overline{B}_0(0,m_P^2,m_P^2)=0$), \ \ \ with
${  \rho_P(s)\equiv \lambda(s,m_P^2,m_P^2)^{\frac{1}{2}}/ q^2 = \sqrt{1-4 m_P^2/s}  }$.

One of the crucial points of the parametrization~\cite{Escribano:2010wt}
employed here is that it incorporates the real part of the logarithm
that comes along with the imaginary part $-i M_\sigma \Gamma_\sigma(s)$
on the basis of analyticity. In the case of narrow-width resonances, these real logs
are essentially negligible and can be dropped. However, if their corresponding
imaginary part is large one naturally expect the appearance of equally large
real logarithms. Moreover, any attempt to match NLO $\chi$PT
at low-energies must incorporate
both the real and imaginary parts of the logs.
Even though our simple approach~\cite{Escribano:2010wt}
can be further refined, it already
contains some of the basic ingredients that makes this matching possible.
Other works that incorporate the real and imaginary parts of the logarithm
in other observables can be found in Refs.~\cite{ND,Sdecays}.

The power behaviour $k=0$ produces an unphysical bound state
in the first Riemann sheet very close below the $\pi\pi$
threshold, which
unnaturally enhanced the amplitude
in the $\eta'\to\eta\pi\pi$~\cite{Escribano:2010wt}, leading in that work to a very small $S\pi\pi$ coupling $c_d=9.9$~MeV.
This case seems to be clearly
disfavoured from the phenomenological point of view and
was
discarded in the analysis of Ref.~\cite{Escribano:2010wt}.
For $k=1$, the amplitude produces just one pole and its correct position
$\sqrt{s^\sigma_{\rm pole}}=  [(441^{+16}_{-\,8})\,-\, i (544^{+18}_{-25})/2] $~MeV~\cite{CCL-sigma}
is recovered for the parameter values $M_\sigma= 806.4$~MeV and $c_\sigma=76.12$.~\footnote{
These are the corresponding central values.
Errors are not discussed in this article. A more detailed numerical analysis
is postponed for a future work. Nonetheless, one may observe that alternative $\sigma$
pole determinations like, e.g.,
$\sqrt{s^\sigma_{\rm pole}}= [(457^{+14}_{-13})\, -\,  i ( 558^{+22}_{-\,14})/2]$~MeV~\cite{GarciaMartin:2011jx},
yield similar central value determinations $M_\sigma=804.1$~MeV and $c_\sigma=70.96$.
This variation gives a preliminary estimate of the expected uncertainties in these quantities.  }
Power behaviours with  $k\geq 2$
are unable to generate the $\sigma$ pole at the right position.
For its closest position, the pole mass is slightly larger and the pole width
is roughly 100 MeV smaller. Likewise, some spurious poles are produced far from
the physical energy range of the problem under study.

For the numerical inputs we will take the $s^k$ scaling with $k=1$
  in Eq.~(\ref{eq.S-self-energy})
and the values $M_\sigma= 806.4$~MeV and $c_\sigma=76.12$.
In these expressions the constants $M_\sigma$ and $c_\sigma$ that appear in the denominator are
parameters set
to agree with the
central value of the  $\sigma$ pole position
$s_\sigma^{\rm pole}= (M_\sigma^{\rm pole}- i\Gamma_\sigma^{\rm pole}/2)^2$
from  Ref.~\cite{CCL-sigma}.

Our estimate of the rescattering of the $\pi\pi$ system related to the isosinglet scalar
is obviously model dependent, as we have introduced an {\it ad hoc}
splitting and self-energy for the scalar multiplet.
The splitting can be easily introduced through the corresponding
terms in the Lagrangian, studied in Ref.~\cite{mass-split}.
On the other hand, while the $1/N_C$ counting would strictly lead to zero-width resonances,
finite widths are needed to regularize the $\tau$ decay phase space integrals
and compare to data. Hence, they need to be taken into account and analyticity requires
the presence of the real logarithm counterparts in the self-energy.
However, if these provide a large contribution,
it seems that $1/N_C$ corrections provide a significant effect
in contradiction with the hypothesis
of neglecting, e.g., resonance-mediated loops.
There is no clear and definitive answer to this issue yet and
one of goals of this work is to explore the raised problem.
In this article, we assume that
this is the only subleading contribution in $1/N_C$ which is numerically relevant
for the current precision of the analysis.
As noticed in Refs.~\cite{RChT-width,RGE},
the resummation of subleading $1/N_C$ corrections can be well defined in perturbation theory
and become crucial even for the $\rho(770)$.
Following previous scalar resonance studies in this line~\cite{Escribano:2010wt},
we consider this resummation of the one-loop $\pi\pi$ self-energy
is also justified, even for the broad $\sigma$:
higher order effects absent in the resummation (multimeson channels)  are completely negligible
below 1~GeV and  the one-loop amplitude seems to provide
the crucial information in our physical range.
Notwithstanding,
this $\pi\pi$ final state interaction
 must be appropriately resummed
in the neighbourhood of the resonance pole,
as noted in Refs.~\cite{RChT-width,RGE}.
Alternatively one might incorporate the $s$--wave rescattering via unitarization
procedures~\cite{Escribano:2010wt,ND} and related dispersion relations
(see, e.g., the semileptonic $B$ decay analysis~\cite{Kang:2013jaa}).
It is important to point out, however, that even in this robust method
only the $\pi\pi$ absorptive corrections are incorporated in the analysis (and the most relevant inelastic intermediate
channels in some cases).

\subsubsection{Incorporating the $f_0$ meson width}

One can  take also into account the $f_0(980)$ width in a similar way. Due the $\sin^2\phi_S$ suppression in~(\ref{eq.MS-splitting}), the $f_0(980)$ produces a clearly
subdominant effect with respect to the impact of the broad~$\sigma$.
The important piece of the self-energy
is its imaginary part, being the real part of its corresponding logarithm
almost negligible in comparison with the leading contribution $M_S^2- s$.
In the case of the narrow $f_0$ resonance,
the location of its pole near the $K\overline{K}$ threshold
will modify  the $f_0$ propagator into the well-known Flatt\'e form~\cite{Flatte:1976xu}
\bear
\Frac{1}{M_{f_0}^2\,-\,s}  \quad \longrightarrow \qquad  \Frac{1}{M_{f_0}^2\,-\,s\, -\, i M_{f_0} \Gamma_{f_0}(s) }  \, ,
\label{eq.f0-propagator}
\eear
with
\bear
M_{f_0} \Gamma_{f_0}(s) &=& \Frac{ c_{f_0}  M_{f_0}^2    \rho_K(s)}{16\pi}\, ,
\eear
which is indeed the near threshold expression of the self-energy
at lowest order in the non-relativistic expansion
in powers of the kaon three-momentum $|\vec{p}_K|\sim \rho_K(s)$~\cite{Braaten:2007dw,Meng:2014ota}.
As the self-energy is only relevant
for $s\approx M_{f_0}^2$, one does not need to consider different
$c_{f_0} s^{k}$ scalings
for the loop corrections as we did for the $\sigma$
meson and the
different values of $k$ amount just for differences
at higher order in the non-relativistic expansion in $\rho_K(s)$.
For $s_{f_0}^{\rm pole} =(M_{f_0}^{\rm pole}-i\Gamma_{f_0}^{\rm  pole}/2)^2
=(990 - i 70/2)^2$~MeV$^2$~\cite{pdg}~\footnote{ We take the central PDG values here.   }
this implies the parameters $M_{f_0}=1024$~MeV and
$c_{f_0} = 17.7$.
The best estimate, based on Roy equations, gives the value
$\sqrt{s_{f_0}^{\rm pole}}=(996^{\,+ 4}_{-14}) - i (24^{+11}_{\,- 3})$~MeV~\cite{Moussallam:2011zg}.
This deviates by less than 1\% from the PDG central value we will use
in Sec.~\ref{sec:Tauola}. We do not expect any difference for our numerical result.
Likewise, in spite of the fact that we have used the average kaon mass $m_K=496$~MeV,
the latter result is not very sensitive to the precise position of the $K\overline{K}$ threshold, with $M_{f_0}$
and $c_{f_0}$
changing by $\pm 0.5\%$ and $\pm 7\%$, respectively, when $m_K$ is varied between the charged and neutral kaon mass values. By far the largest
effect would be the uncertainty in the $f_0$ mass and width with
errors
of $\pm 20$~MeV and $\pm 30$~MeV, respectively~\cite{pdg}.

Therefore, for the numerical inputs we will take
 $M_{f_0}=1024$~MeV
and $c_{f_0} = 17.7$.~\footnote{
We remind that the parameter $M_{f_0}$ is not the pole mass $M_{f_0}^{\rm pole}$.}

\section{The decay $\tau \to \pi\pi\pi \nu_\tau$ through tensor resonances}
\label{sec:T}

In this section we focus on   tau decay into three pions through an intermediate tensor resonance
 ($J^{PC}=2^{++}$)  in the cascade decay $\tau\, \to \,\nu_\tau \, \pi^- \,T(\to \pi\pi)$.
Our study reproduces the prediction for the tau decay into a tensor resonance and
a chiral pseudo-Goldstone~\cite{Castro:2011zd}
and expands then for the case of the off-shell tensor resonance.

$G$-parity conservation implies that for the non-strange axial-vector current
(with $G=-1$)
the tensor resonance produced in combination with a pion
must have $G=(-1)^I=+1$ and, hence, even isospin.
As a consequence of this, it must
be an isosinglet in the case of $q\bar{q}$ multiplets
($T=f_2(1270)$, $f_2(1430)$, $f_2'(1525),\, f_2(1565)...$). In this article we study the impact of the lightest
tensor, $f_2(1270)$, which dominantly decays into $\pi\pi$~\cite{pdg}.
The $f_2'(1525)$ mainly goes into $K\overline{K}$ and
has a negligible decay into $\pi\pi$~\cite{pdg}.
Our analysis is then restricted to the lowest tensor resonances.
We discarded not so well established resonances
such as the $f_2(1430)$ and $f_2(1565)$, whose $\pi\pi$ partial width
are not determined in any of the references quoted by PDG~\cite{pdg}.
In addition, we would like to stress that,
the contribution from the $f_2(1270)$
is found to be highly suppressed in our later numerical analysis,
as it is placed near the $\pi^0\pi^0$ spectrum end point
(or the $\pi^+\pi^-$ spectrum
for $\tau\to\nu_\tau \pi^-\pi^-\pi^+$),
$M_{\pi\pi}^{\rm end}=M_\tau-m_{\pi^\pm}\simeq 1637$~MeV.
Thus, heavier $f_2$ resonances should have even stronger phase-space
suppressions. In particular the $f_2(1640)$ and further tensors
lie beyond $M_{\pi\pi}^{\rm end}$.

\subsection{The R$\chi$T Lagrangian for tensor fields}

\begin{figure}[!t]
\begin{center}
\includegraphics[width = 0.4\textwidth]{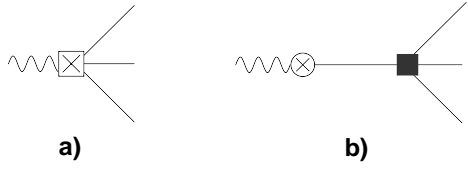}
\end{center}
\caption{{\small
 New diagrams due to the short-distance
    $\cO(p^4)$ operators $L_{1,2,3}^{SD}$.  For a more detailed explanation, see the text.
  The vertices from $\mL_{\rm non-R}^{(4)}$ ($\mL_{\rm non-R}^{(2)}$)
  are represented by squares (circles).
  The straight lines are pions and the wavy ones correspond to the incoming $W^-$.
}}
\label{fig.diagr-nonR}
\end{figure}

The relevant part of the chiral invariant Lagrangian for the pion-tensor
production (Fig~\ref{fig.diagr}) consists in this case of
\begin{itemize}
\item Operators with one resonance field~\cite{rcht,Zauner:2007},~\footnote{
 There are two more operators for $\mL_T$ in Ref.~\cite{Zauner:2007}
allowed by chiral symmetry
but they contain the trace $T_{\,\, \alpha}^\alpha$~\cite{Zauner:2007}:
$\Delta \mL_T|_{ \mbox{\tiny off-shell} }=
\bra T_{\,\, \alpha}^\alpha \left(  \beta u^\mu u_\mu +\gamma \chi_+ \right) \ket$.
Since they are proportional to the equations of motion
of the tensor, which on-shell require it to be transverse ($\nabla^\alpha T_{\alpha\beta}=0$)
and traceless ($T_{\,\,\, \alpha}^\alpha=0$), they can be removed through meson field redefinitions
and we will not discuss them in the present work.
}

\bear
\mL_A &=& \Frac{F_A}{2\sqrt{2}} \bra A_{\mu\nu} f_-^{\mu\nu} \ket\, ,
\nn\\
\mL_T &=& g_T\bra T_{\mu\nu} \{ u^\mu,u^\nu\}\ket\, .
\label{eq.LT}
\eear
\item Operators with an axial-vector and a tensor field
(which provides the $AT\pi$ vertex in
diagram c) in Fig.~\ref{fig.diagr}),
\bear
\mL_{AT\pi} &=&
 \lambda_1^{AT}\, \bra\{T_{\mu\nu}, A^{\nu\alpha} \} h_\alpha^\mu\ket
 + \lambda_2^{AT} \bra\{ A_{\alpha\beta} , \nabla^\alpha T^{\mu\beta} \} u_\mu\ket
\, , \label{eq.tensor-lagr}
\eear
with $h_{\alpha\mu}=\nabla_\alpha u_\mu +\nabla_\mu u_\alpha$~\cite{rcht}.
Only the independent operators from $\mL_{AT}$ that contribute to the $AT\pi$
vertex are shown here. We construct here the general
chiral invariant operators at lowest order in derivatives, $\cO(p^2)$,
that may contribute to the $AT\pi$ vertex.~\footnote{
There are also two more $AT\pi$ operators allowed by symmetry
but they contain the trace $T_{\,\, \alpha}^\alpha$ or the contraction
$\nabla^{\alpha} T_{\alpha\beta}$:
$\Delta \mL_{AT\pi}|_{ \mbox{\tiny off-shell} } =
\beta_{AT\pi} \bra \{ A_{\alpha\beta} , \nabla^\alpha T_{\,\, \mu}^\mu \}  u^\beta \ket
+ \gamma_{AT\pi} \bra \{ A_{\alpha\beta} , \nabla_\mu T^{\mu\alpha}\} u^\beta \ket$.
They do not propagate the tensor meson and can be removed from the generating functional
through appropriate field redefinitions.
}
\item Operators without resonance fields~\cite{Zauner:2007}:
 in addition to~(\ref{eq.non-R-lagr}) we have
\bear
 \mL_{\rm non-R}^{(4)}    &=&
 L_1^{SD}\bra u^\mu u_\mu \ket ^2\,
\,+\,  L_2^{SD} \bra u^\mu u^\nu \ket \,\bra u_\mu u_\nu \ket
\, + \,  L_3^{SD}\bra (u^\mu u_\mu)^2 \ket \, ,
\label{eq.non-R-lagr+Op4}
\eear
with~\cite{Zauner:2007}

\bear
L_2^{SD}=2 L_1^{SD}= - \Frac{L_3^{SD}}{2} = -\Frac{g_T^2}{M_T^2}\, .
\label{eq.LjSD}
\eear
The appearance of $\mL_{\rm non-R}^{(4)} $ was explained in ~\cite{Zauner:2007}:
in order to reproduce the correct short-distance behaviour
for the forward $\pi\pi$ scattering
--prescribed by the Froissart bound~\cite{Froissart:1961ux}--
one must add  non-resonant
$\cO(p^4)$ terms with appropriate $L_{1,2,3}^{SD}$.
As a consequence this,  new non-resonant diagrams  generated by $L_{1,2,3}^{SD}$
(Fig.~\ref{fig.diagr-nonR}) have to be included in the calculation of the $3\pi$-AFF.
Additional details from Ref.~\cite{Zauner:2007} are provided in App.~\ref{app.ChPT}.
This problem did not appear in the scalar and vector resonance case~\cite{rcht},
i.e. the introduction of the scalar and vector resonance
interaction,  $\mL_S$ and $\mL_V$~\cite{Dumm:2009va},  did not spoil the high-energy behaviour
of the forward pion scattering and no additional $\cO(p^4)$ terms were required~\cite{rcht}.

\end{itemize}

We will assume the ideal mixing in the tensor nonet
$T_{\mu\nu}=T^a_{\mu\nu} \lambda^a/\sqrt{2}$
and that the $f_2(1270)$ resonance is the pure
$u\bar{u}+d\bar{d}$ component:
\bear
T^{\mu\nu} &=& \left(\begin{array}{ccc}
\Frac{f_2^{\mu\nu}}{\sqrt{2}} &0&0 \\ 0 & \Frac{f_2^{\mu\nu}}{\sqrt{2}} & 0 \\
0&0&0
  \end{array}\right)\quad +\quad ...
\eear

\subsection{AFF into $T\pi^-$}
\label{sec:AFF}

The general possible structure for the hadronic matrix element into a tensor and a pion is given by
three independent form-factors~\cite{Castro:2011zd}, which can be arranged in the form
\bear
\bra f_2(k,\epsilon)\, \pi^-( p_3 )|\,\bar{d}\gamma^\alpha\gamma_5 u\, |0\ket &=&
 \epsilon_{\mu\nu}^* H^{\alpha,\, \mu\nu}_{  T \pi}
\label{eq.general-f2pi-AFF}
\\
&& \hspace*{-5.cm}
= i \,\epsilon_{\mu\nu}^* \,
\left[  P_T(q)^{\alpha\rho} \, p_3^\nu
\, \left(\, g_\rho^\mu \, \mF^a_{T\pi}(q^2 ; k^2 ) \,\,  +\,\,
 p_{3\, \rho} p_3^\mu \,\mG^a_{T\pi}(q^2; k^2)
\,\right)
\,\, +\,\,
 p_3^\mu p_3^\nu q^\alpha \, \mH^a_{T\pi}(q^2; k^2)
\right]\, ,
\nn
\eear
with $q= p_3 +k$ and $\epsilon_{\mu\nu}$ the polarization of the outgoing tensor~\cite{Castro:2011zd,Zauner:2007}.
Due to the partial conservation of the axial-vector current, the
$\mH^a_{T\pi}(q^2;k^2)$ form-factor is suppressed by $m_\pi^2$.

Here the tensor resonance has been assumed to be the asymptotic final state with polarizations fulfilling the on-shell constraints~\cite{Zauner:2007}
\bear
\epsilon_{\mu\nu}=\epsilon_{\nu\mu}\, , \qquad k^\mu \epsilon_{\mu\nu}=0\, ,
\qquad g^{\mu\nu} \epsilon_{\mu\nu}=0\, .
\eear

We used the completeness relation~\cite{Zauner:2007,LopezCastro:1997im}
\bear
\!\!\!\!\!\!\!\!\mP(k)^{\mu\nu,\alpha\beta}=\sum_{\epsilon}\epsilon_{\mu\nu}\epsilon^*_{\alpha\beta}
&=&\Frac{1}{2}\left(P(k)^{\mu\alpha}P(k)^{\nu\beta}+P(k)^{\nu\alpha}P(k)^{\mu\beta}\right)-\Frac{1}{3}P(k)^{\mu\nu}P(k)^{\alpha\beta} \,\,
\label{eq.tens-deno}
\eear
with $P(k)_{\mu\nu}= \left.P_T(k)^{\mu\nu}\right|_{k^2=M_T^2} =g_{\mu\nu}
-k_\mu k_\nu/M_{T}^2$.

The hadronic Lagrangian from Eqs.~(\ref{eq.LT}) and~(\ref{eq.tensor-lagr}) leads to the
determination
\bear
\mF^a_{T\pi}(q^2  ;k^2   ) &=& - \Frac{8g_T}{ F_\pi  }
  +  \Frac{4\sqrt{2} F_A\lambda_1^{AT}}{F_\pi} \Frac{  (q p_3)    }{M_A^2-q^2}
-  \Frac{2\sqrt{2} F_A\lambda_2^{AT}}{F_\pi} \Frac{  (qk)    }{M_A^2-q^2}
\, ,
\nn\\
\mG^a_{T\pi}(q^2 ;k^2  ) &=&
- \Frac{4\sqrt{2} F_A\lambda_1^{AT}}{ F_\pi  } \Frac{1  }{M_A^2-q^2}
 -  \Frac{2\sqrt{2} F_A\lambda_2^{AT}}{  F_\pi  } \Frac{1  }{M_A^2-q^2}
\, ,
\nn\\
\mH^a_{T\pi}(q^2 ;k^2  ) &=& 0 \, ,
\eear
with $(qp_3)=(q^2+m_\pi^2-k^2)/2$ and $(qk)=(q^2-m_\pi^2+k^2)$. Even
though $k^2=M_T^2$ when the tensor resonance is on-shell we have kept the off-shell momentum dependence
stemming from our R$\chi$T Lagrangian.
The $m_\pi^2$ chiral suppressed form-factor $\mH^a_{T\pi}(q^2)$
is exactly zero in our approach as we are considering
a resonance Lagrangian with the lowest number of derivatives
(this is, two derivatives, $\cO(p^2)$)
and the Lorentz structure corresponding to $\mH^a_{T\pi}(q^2;k^2)$ carries three powers of external momenta.

If one imposes a vanishing behaviour for the contribution of the $T\pi$ absorptive cut to the
axial-vector correlator at $q^2\to \infty$ one finds that the form-factors vanish at large momentum transfer like
$\mF^a_{T\pi}(q^2;M_T^2)\stackrel{q^2\to \infty}{\longrightarrow} \cO(1/q^2)$ and
$\mG^a_{T\pi}(q^2;M_T^2)\stackrel{q^2\to \infty}{\longrightarrow} \cO(1/q^4)$ or faster
(see App.~\ref{app.optical-theorem} for details).
Demanding this to the previous R$\chi$T form-factors
$\mF^a_{T\pi}$ and $\mG^a_{T\pi}$
yields, respectively, the constraints (taking into account $k^2 = M_T^2$
for the on-shell resonance),
\bear
4 \sqrt{2} g_T \, +\,  2 F_A\lambda_1^{AT}  \, -\, F_A\lambda_2^{AT} &=& 0\, ,
\qquad\qquad
2\lambda_1^{AT} \, +\, \lambda_2^{AT} \,=\, 0 \, .
\label{eq.constraints1}
\eear
This leads to the resonance coupling relations
\bear
F_A\lambda_2^{AT}  &=&  - 2 F_A\lambda_1^{AT} \,=\, 2\sqrt{2} g_T\,
\label{eq.constraints2}
\eear
and the form-factors
\bear
\mF^a_{T\pi}(q^2  ;k^2   ) &=& - \Frac{8g_T}{  F_\pi   }  \Frac{ M_A^2  }{M_A^2-q^2}\, ,
\nn\\
\mG^a_{T\pi}(q^2  ;k^2   ) &=&    0\, .
\label{eq.TP-AFF+constraints}
\eear

This result agrees with that in Ref.~\cite{Castro:2011zd} near the axial-vector resonance.
Furthermore, in the chiral limit, if one requires the same fall-off for the form-factors
therein
one has an agreement in the full energy range.
Additional details can be found in App.~\ref{app.Castro-comparison}.

\subsection{$3\pi$ AFF through an intermediate tensor resonance}
\label{sec:3pi-from-T}

The three possible decay  mechanisms involving the tensor resonance are drawn in Fig.~\ref{fig.diagr}.
We present here some useful intermediate results.

The $\pi^0\pi^0\pi^-$ production with the neutral pions mediated by a tensor resonance is provided
by three ingredients:
\begin{itemize}
\item
The transition $W^{-\mu}(q) \to f_2(k)^* \pi^0(p_3)$ taking into account the three diagrams
is given by
\bear
&&\bra  f_2^*(k,\epsilon)
\pi^- (p_3) | \bar{d}\gamma^\mu \gamma_5 u|0\ket \,=\, \epsilon_{\alpha\beta}^{*} H^{\mu,\, \alpha\beta}_{T\pi}
\label{eq.tens_axvec_pion}
\\
&&\qquad=\,
\frac{-4\sqrt{2}\, i}{F_\pi }p_{3}^\alpha\epsilon^{\star \alpha\beta}\Bigl[
\sqrt{2}g_T \left(g_{\beta\mu} -\frac{q_\beta q_\mu}{q^2 -m_\pi^2}\right)
\nn\\
&&\qquad\qquad\qquad\qquad\qquad
 -
F_A  \frac{\left[
\lambda_1^{AT} (qp_3 g_{\beta\mu} - q_\beta p_{3\mu})
\, -\,\frac{1}{2}
\lambda_2^{AT} (qk g_{\beta\mu} - q_\beta k_{\mu})
 \right] }{M_A^2 -q^2}
\Bigr]
 \, .
\nn
\eear
   After imposing the high-energy constraints~(\ref{eq.constraints2}), this expression gets greatly simplified into
\bear
H^{\mu,\, \alpha\beta}_{T\pi}
&=&
\frac{-8\, i g_T}{ F_\pi }p_{3}^\alpha
\Bigl[  \frac{M_A^2 }{M_A^2 -q^2} \, P_T(q)^{\beta\mu}
\, -\,
\Frac{m_\pi^2 q_\beta q_\mu}{q^2 (q^2-m_\pi^2)}
\Bigr]
 \, .
\label{eq.tens_axvec_pion+SD}
\eear
We remark that we have not used the on-shell conditions in Eqs.~(\ref{eq.tens_axvec_pion}) and (\ref{eq.tens_axvec_pion+SD})
above.
\item The tensor propagator~\cite{Zauner:2007}:
\bear
\Delta_T(k)^{\mu\nu,\alpha\beta}&=& \Frac{i\, \mP(k)^{\mu\nu,\alpha\beta}}{M_T^2-k^2}\, .
\eear

\item The decay amplitude $\mM(   f_2^*(k)  \to \pi^0(p_1)\pi^0(p_2))= \epsilon^{\alpha\beta}\Gamma_{\alpha\beta} $ is given by
  \bear
\Gamma_{\alpha\beta} &=&
\frac{-i \sqrt{2} g_T}{F_\pi^2} \Bigl[
k^\alpha k^\beta
- \Delta p^\alpha\, \Delta p^\beta
\Bigr]    \, ,     \label{eq.tens_two_pions}
\eear
with $\Delta p^\rho=p_1^\rho-p_2^\rho$ and $k^2=s_3$.
No on-shell condition has been assumed in the expression above.
The term $k^\alpha k^\beta$ becomes zero when contracted with the $\epsilon^{\alpha\beta}$
polarization of an external on-shell tensor resonance.

\end{itemize}

The $\pi^0\pi^0\pi^-$ AFF is then given by
\bear
H^\mu &=&
H_{(0)}^\mu \, +\,
H_{T\pi}(k,p_3)^{\mu,\, \alpha\beta} \,\,\, \Delta_T(k)_{\alpha\beta,\rho\sigma} \,\,\, \Gamma(p_1,p_2)^{\rho\sigma}
\label{eq.Htens}
\\
&&\,\,\,=\,\,\,
H^\mu_{(0)}\,\,\,+ \,\,\,
H^\mu_{(1)} + \frac{H^\mu_{(2)}}{M_{T}^2 - s_3}
\, .  \nn
\eear
The first term, $H_{(0)}^\mu$, comes from the non-resonant diagrams in Fig.~\ref{fig.diagr-nonR} generated
by the short-distance terms $L_{1,2,3}^{SD}$ in Eqs.~(\ref{eq.non-R-lagr+Op4}) and (\ref{eq.LjSD}). The
second and third ones, $H_{(1)}^\mu$ and $H_{(2)}^\mu$, respectively, are produced by the
diagrams with tensor resonance exchanges (Fig.~\ref{fig.diagr}).
$H_{(1)}^\mu$  comes from the $k^\alpha k^\beta$ term in the
$\Gamma[T(k)_{\alpha\beta} \to\pi^0(p_1)\pi^0(p_2)]$ vertex function
and does not contribute to the on-shell decay $T\to\pi^0\pi^0$.
For sake of this,
the contribution with $H_{(1)}^\mu$ does not propagate the tensor resonance
and has no pole at $s_3=M_T^2$.
The contribution to the three-pion AFF from the remaining part of the $T\pi^0\pi^0$ vertex is encoded
in $H_{(2)}^\mu$.

The value of these two types of contributions are
\bear
 H^\mu_{(0)} &=& \Frac{8\sqrt{2} i g_T^2}{3F_\pi^3 M_T^2} P_T(q)^{\mu\nu}
\big[(s_3 - s_2 + 2s_1   -4 m_\pi^2 )(p_1 - p_3)_\nu
\nn\\
&& \qquad\qquad\qquad \qquad\qquad
  + (s_3 - s_1 + 2s_2 -4 m_\pi^2)(p_2 - p_3)_\nu \big]
\label{eq.contri0}
\\
 &&
 - \Frac{8\sqrt{2} i g_T^2 m_\pi^2}{F_\pi^3 M_T^2 q^2 (q^2-m_\pi^2) }
  \, q^\mu\,   \left( s_1 s_2  - m_\pi^2 q^2 - m_\pi^4\right)
  \nn\\
  && \nn \\
  && \nn \\
 H^\mu_{(1)} &=& \Frac{8\sqrt{2} i g_T^2}{F_\pi^3 M_T^2}
\Frac{m_\pi^2}{q^2(q^2-m_\pi^2)}
\,  q^\mu\, \left[ (k q) (k p_3) - \frac{s_3}{3} \left( (qp_3)
+ \Frac{ 2 (k q) (k p_3) }{M_T^2} \right)\right]
\label{eq.contri1}
\\
&&
-\, \Frac{8 i g_T}{F_\pi^3 M_T^2}  \Frac{M_A^2}{(M_A^2-q^2)} P_T(q)^{\mu\nu} k_\nu
\bigg[
\sqrt{2} g_T
\left(   \left(1 -\Frac{2 s_3}{3 M_T^2} \right)
   (kp_3) + \Frac{s_3}{3}  \right)
\nn\\
&&
\qquad +    (F_A\lambda_1^{AT} +\sqrt{2} g_T) \Frac{ q^2 (kp_3)}{M_A^2} \,
  \left(\Frac{2 s_3}{3 M_T^2} -1\right)
+   (F_A\lambda_2^{AT} -2\sqrt{2} g_T) \Frac{ q^2 s_3 }{  6 M_A^2}
\bigg]\, ,
\nn
\\
&& \nn \\
&& \nn
\\
H^\mu_{(2)\,{\rm a_1-pole}} &=&
-\, \Frac{8 i g_T}{F_\pi^3}
\Frac{F_A}{M_A^2-q^2} P_T(q)^{\mu\nu}
\bigg[
\left( \lambda_1^{AT} M_A^2 - \left(\lambda_1^{AT}+\Frac{\lambda_2^{AT}}{2} \right) (kq)      \right)
\, (q\Delta p)\, \Delta p_\nu
\nn\\
&&
\hspace*{-1.5cm}
+ \left( \Frac{\lambda_1^{AT} M_A^2 (\Delta p)^2
( kp_3 + M_T^2)}{3 M_T^2}
+  \left(\lambda_1^{AT}+\Frac{\lambda_2^{AT}}{2} \right)
\left(  (q\Delta p)^2 -  \Frac{(\Delta p)^2   M_A^2}{3} \right)\right) k^\nu \bigg]  \, ,
\label{eq.contri2A}
\\
&& \nn
\\
  && \nn \\
H^\mu_{(2)\,{\rm a_1\, no-pole}} &=&
-\, \Frac{2\sqrt{2}  i g_T}{F_\pi^3} P_T(q)^{\mu\nu}
\bigg[
- 2\sqrt{2}(F_A\lambda_1^{AT} +\sqrt{2} g_T)
\left( (q\Delta p) \Delta p_\nu + \Frac{(kp_3) (\Delta p)^2 }{3 M_T^2} k_\nu\right)
\nn
\\
&&
+ \sqrt{2}( F_A\lambda_2^{AT} -2\sqrt{2} g_T   ) \Frac{(\Delta p)^2 }{3 } k_\nu
 \bigg]
\nn\\
&&
-\, \Frac{8\sqrt{2}  i g_T^2 m_\pi^2}{  F_\pi^3 q^2 (q^2-m_\pi^2)} \, q^\mu\,
\bigg[ (q\Delta p)^2  + \Frac{(\Delta p)^2}{3 M_T^2}
\left(kq\, kp_3 -qp_3 M_T^2\right)
\bigg]\, ,
\label{eq.contri2B}
\eear
with $(\Delta p)^2= 4 m_\pi^2-s_3$,
$(kq)=(q^2+s_3-m_\pi^2)/2$ and
$(k\Delta p) =0$. From these, one can derive a series of dependent scalars:
$(kp_3)=(qk)-s_3=(q^2-s_3-m_\pi^2)/2$,
$(qp_3)=q^2-(qk)=(q^2-s_3+m_\pi^2)/2$, $(q\Delta p)=(p_3\Delta p)= (s_2-s_1)/2$
and the relation $s_{1,2}= kp_3 + 2m_\pi^2 \mp q\Delta p$.
For convenience we have split $H^\mu_{(2)}$
into its parts with and without the $a_1$ pole.
We also used the relation
$ (q p_3) k^\mu - (q k) p_3^\mu= q^2 P_T(q)^{\mu\nu} k_\nu$.

We now combine $H_{(0)}^\mu$, $H_{(1)}^\mu$ and $H_{(2)}^\mu$ and rewrite
their sum in terms of the Lorentz decomposition~(\ref{eq.hadr-curr-3pions}).
This provides the contribution to the ${\pi^0\pi^0\pi^-}$ AFFs in~(\ref{eq.hadr-curr-3pions})
derived from tensor resonance exchanges:
\bear
\mF_1^{\pi^0\pi^0\pi^-}(s_1,s_2,q^2) \bigg|_T
&=&
\mF_{1,\,\, (0)}^{\pi^0\pi^0\pi^-} (s_1,s_2,q^2)  \, +\,
\mF^{\pi^0\pi^0\pi^-}_{1,\,\,\rm (RSD)} (s_1,s_2,q^2)
\label{eq.ff-tens}
\\
  && \nn \\
&&   - \Frac{4}{9  F_\pi^3 }\Frac{  g_T}{M_T^2}
\Frac{( F_A\lambda_2^{AT} - 2\sqrt{2} g_T)}{M_A^2-q^2}
\times \bigg[
s_3 q^2
\nn\\
&&\qquad  + \Frac{M_T^2}{M_T^2-s_3} \left(
 3 (q\Delta p)^2
  - 9 (qk) (q\Delta p)  - q^2 (\Delta p)^2
\right)
\bigg]
\nn\\
&&  - \Frac{8}{3 F_\pi^3}    \Frac{g_T}{M_T^2}
\Frac{(F_A\lambda_1^{AT} + \sqrt{2} g_T)}{M_A^2-q^2}
  \times
\bigg[
q^2 (kp_3) \left(\Frac{2 s_3}{3 M_T^2} -1\right)
\nn\\
&&\qquad
+\Frac{M_T^2}{M_T^2-s_3}
\left( (q\Delta p)^2 + 3 (q\Delta p) (qp_3)+ \Frac{ q^2 (kp_3) (\Delta p)^2}{3M_T^2} \right)
\bigg]
\nn\\
\mF_P^{\pi^0\pi^0\pi^-}(s_1,s_2,q^2) \bigg|_T &=&
\mF_{P,\,\, (0)}^{\pi^0\pi^0\pi^-} (s_1,s_2,q^2)  \,
\label{eq.ff-tensP}
\\ &&
+
 \frac{8 \sqrt{2}g_T^2 m_\pi^2}{3 M_T^2 F_\pi^3 q^2 (m_\pi^2 -q^2)}
\times \bigg[
(qp_3) s_3  +   (kq) (kp_3) \left( \Frac{2 s_3}{M_T^2}   -3 \right)
\nn\\
&&
\qquad +\Frac{M_T^2}{M_T^2-s_3} \left(
3 (q\Delta p)^2 + \left(\Frac{(kq) (kp_3)}{M_T^2} - (qp_3)  \right)(\Delta p)^2\right)
\Bigg] \, ,
\nn
\eear
with
\bear
  \mF_{1,\,\, (0)}^{\pi^0\pi^0\pi^-}   (s_1,s_2,q^2)
&=&
  \Frac{8\sqrt{2} g_T^2}{3 F_\pi^3 M_T^2}  (
2 s_1 -s_2+s_3
-4 m_\pi^2) \, ,
\label{eq.F1-0}
\\  && \nn \\
 \mF_{P,\,\, (0)}^{\pi^0\pi^0\pi^-}   (s_1,s_2,q^2)   &=&
    - \Frac{8\sqrt{2}  g_T^2  m_\pi^2}{ F_\pi^3 M_T^2 q^2 (q^2-m_\pi^2) }
    \left( s_1 s_2  - m_\pi^2 q^2 - m_\pi^4\right) \, ,
\label{eq.FP-0}
\\  && \nn \\
\mF^{\pi^0\pi^0\pi^-}_{1,\,\, \rm  (RSD)} (s_1,s_2,q^2)
&=&
- \Frac{8 \sqrt{2}  }{3 F_\pi^3}\Frac{g_T^2}{ M_T^2} \Frac{M_A^2}{M_A^2-q^2}
\bigg[
(kp_3) + \Frac{s_3}{3}\left(1-\Frac{2 (kp_3)}{M_T^2}\right)
\nn\\
&&\qquad
- \Frac{M_T^2}{M_T^2-s_3}\left(
3 (q\Delta p) + \Frac{(\Delta p)^2}{3}
+ \Frac{(kp_3)(\Delta p)^2 }{3 M_T^2}
\right)
\bigg] \, \label{eq.tens_rsd},
\eear
where the contributions $\mF_{1,\,\, (0)}^{\pi^0\pi^0\pi^-}$
and $\mF_{P,\,\, (0)}^{\pi^0\pi^0\pi^-}$
come from the $H_{(0)}^\mu$ part of the matrix element $H^\mu$.

All the results here refer to the $\pi^0\pi^0\pi^-$ AFF.
Isospin symmetry~\cite{Pais:1960zz,Finkemeier:1996,Girlanda:1999fu}
relates them to the $\pi^-\pi^-\pi^+$ form-factors,
which can be obtained by mean of the relations~(\ref{eq.isospin-rel2}).

The expression of the form-factors get greatly simplified
after applying the high-energy constraints
extracted from the analysis of the $T\pi$ AFF in Eq.~(\ref{eq.constraints2}):
\bear
\mF_1^{\pi^0\pi^0\pi^-}(s_1,s_2,q^2) \bigg|_T&=&
\mF_{1,\,\, (0)}^{\pi^0\pi^0\pi^-}(s_1,s_2,q^2)
\, +\,
\mF^{\pi^0\pi^0\pi^-}_{1,\,\, \rm  (RSD)}(s_1,s_2,q^2)
\, ,
\eear
while these resonance short-distance conditions do not affect
the longitudinal form-factor $\mF_P(s_1,s_2,q^2) \bigg|_T$, which remains the same as
in~(\ref{eq.ff-tensP}).

The comparison between CLEO's results and ours for the amplitude and the related AFF is given in App.~\ref{app.cleo}. From that, we conclude that the two parametrizations coincide near the resonance energy regions ($s_3 \simeq M_T^2$, $q^2 \simeq M_A^2$). However,
for an arbitrary off-shell momentum
we have a more general momentum structure which ensures
the right low energy behaviour and the transversality of the matrix element in the chiral limit,
allowing a proper matching with $\chi$PT.

\subsection{Tensor resonance width}
\label{sec:tensor-width}

In order to include the effect of the tensor width,
we modify the tensor resonance propagator in the form
\bear
\frac{1}{M_T^2 - s}  \qquad \longrightarrow \qquad \frac{1}{M_{f_2}^2 - s - i M_{f_2} \Gamma_{f_2}(s)}\, ,
\eear
with the spin--2 energy-dependent Breit-Wigner width used in CLEO's analysis~\cite{CLEO:1999},
\bear
\Gamma_{f_2}(s)  &=&  \, \Gamma^{f_2}_{0} \, \Frac{ s^2 }{M_{f_2}^4}
\, \Frac{\rho_\pi(s)^5}{\rho_\pi(M_{f_2}^2)^{\,5}  } \,\, .
\eear
For the numerical estimation in the next Section we will take
the PDG central value $\Gamma^{f_2}_0 = 186.7$~MeV
for the $f_2(1270)$ total decay width~\cite{pdg}.

The tensor contribution to the AFF depends on the $g_T$ coupling, which is related to the on-shell
decay width into two pseudo-Goldstones~\cite{Zauner:2007}:
\bear
\Gamma_{f_2\to \pi\pi}  &=&\Frac{g_T^2 M_{f_2}^3 \rho_{\pi}(M_{f_2}^2)^{\,5}}{  40\pi F_\pi^4}\, .
\eear
Using the PDG central values,  $\Gamma_{f_2\to\pi\pi}^{\rm exp} =157.2$~MeV, $M_{f_2} =1275.5$~MeV,
$m_\pi=139.57$~MeV and  $F_\pi=92.2$~MeV, one obtains
\bear\label{eq.value-gt}
g_T \simeq 28\, \mbox{MeV}\, ,
\eear
which agrees with the estimation in~\cite{Zauner:2007}.

\section{Implementation in Tauola: numerical results}
\label{sec:Tauola}

In the previous sections we described the set
of the three pion form
factor contributions related with the tensor and scalar
intermediate resonances and calculated on the base of the R$\chi$T Lagrangians.
In this section we present
a first numerical estimate with
the updated version of the Monte Carlo (MC) event generator Tauola~\cite{Jadach:1993hs}.
It incorporates
the new scalar and tensor contributions to the AFF computed in this article,
provided in~(\ref{eq.ff-scal}) and (\ref{eq.ff-tens}), respectively.
\footnote{The MC Tauola implementation of these channels was cross-checked with a Mathematica code, which can be provided on demand. }

\begin{figure}[!t]
\begin{center}
\includegraphics[width = 0.45\textwidth, height = 0.35\textwidth]{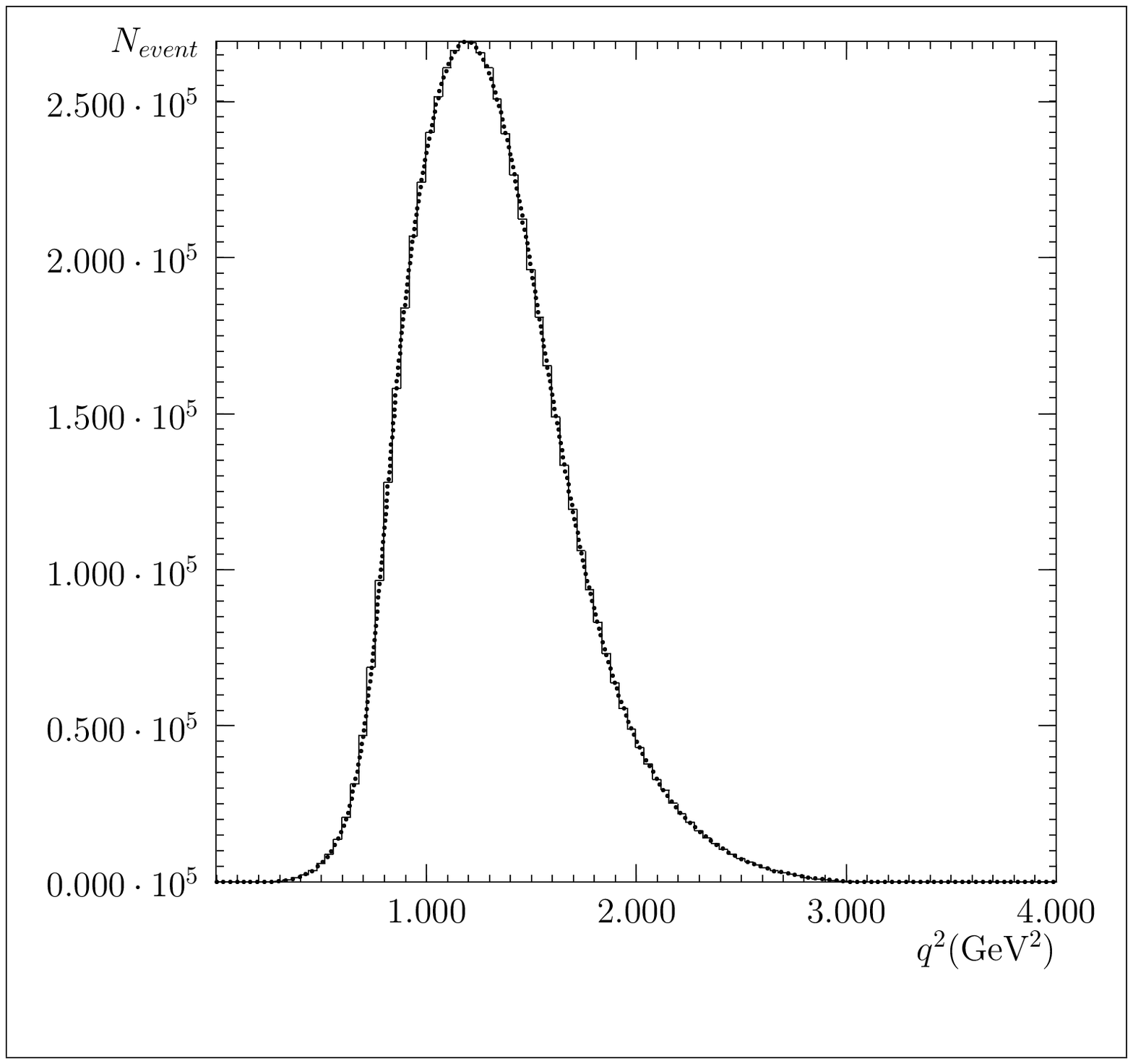}
\includegraphics[width = 0.45\textwidth, height = 0.35\textwidth]{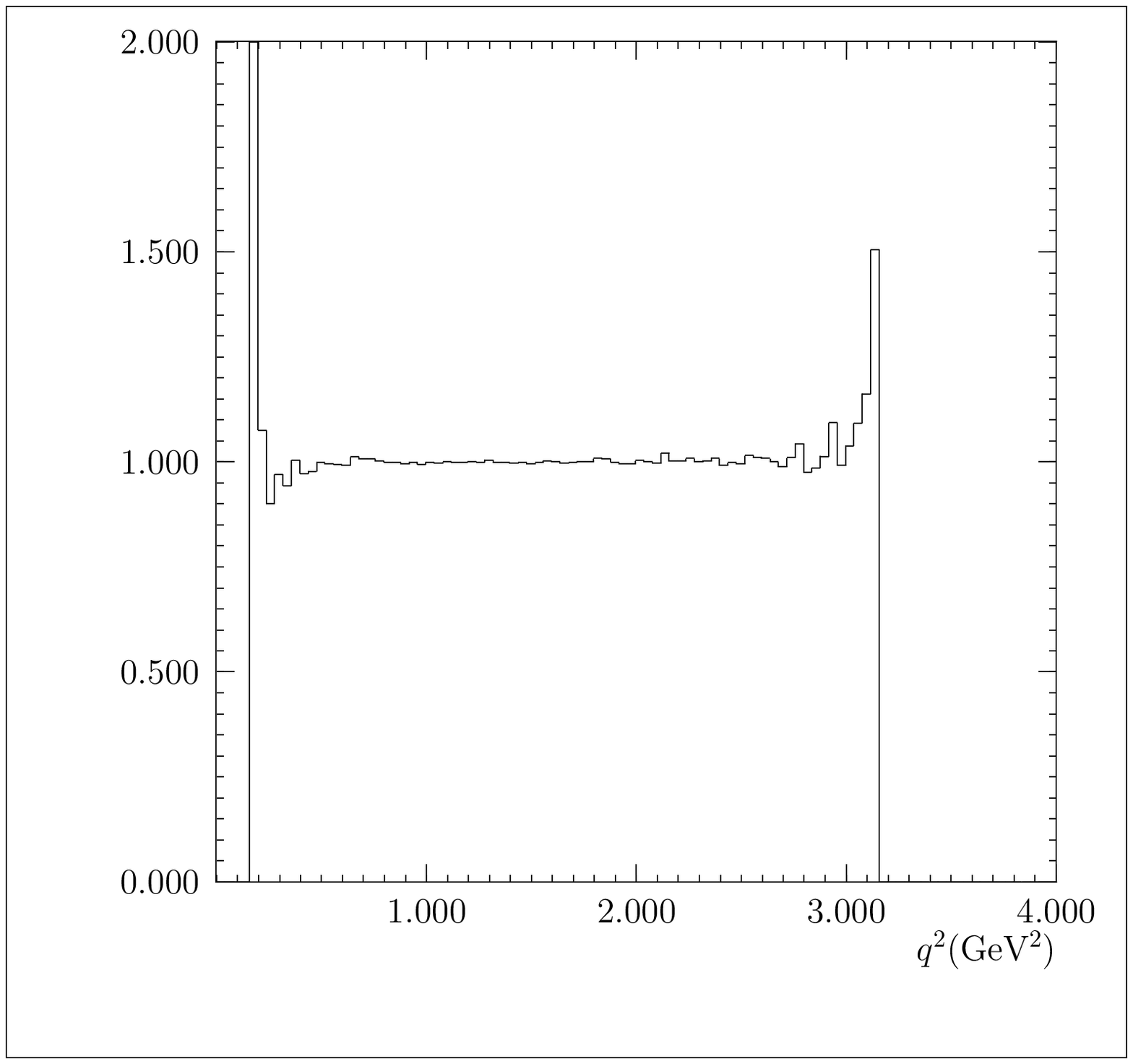}
\end{center}
\caption{{\small
 Three pion $q^2$ spectrum $d\Gamma^{\pi^0\pi^0\pi^-}/dq^2$
(left) and the ratio
of the MC and the analytical $q^2$ spectrum (right). }
}
\label{fig.tauola_compar}
\end{figure}

First, we compare the analytical and Tauola distributions
for the decay width ($d\Gamma^{\pi\pi\pi}/dq^2$)
and repeat the tests on numerical stability
of the MC, as in Sec.~4 of Ref.~\cite{Shekhovtsova:2012ra}~\footnote{
We use the same samples and integration procedure
as in~\cite{Shekhovtsova:2012ra}.
The MC result here corresponds to a number of events $N_{\rm ev.}=6\cdot 10^6$.}
For further details see this reference.
The  comparison is presented in Fig.~\ref{fig.tauola_compar}.
We present here only  $d\Gamma^{\pi^0\pi^0\pi^-}/dq^2$ spectrum. A
similar result has been obtained for the $\pi^-\pi^-\pi^+$ mode.

\begin{figure}[!t]
\begin{center}
\includegraphics[width = 1.\textwidth]{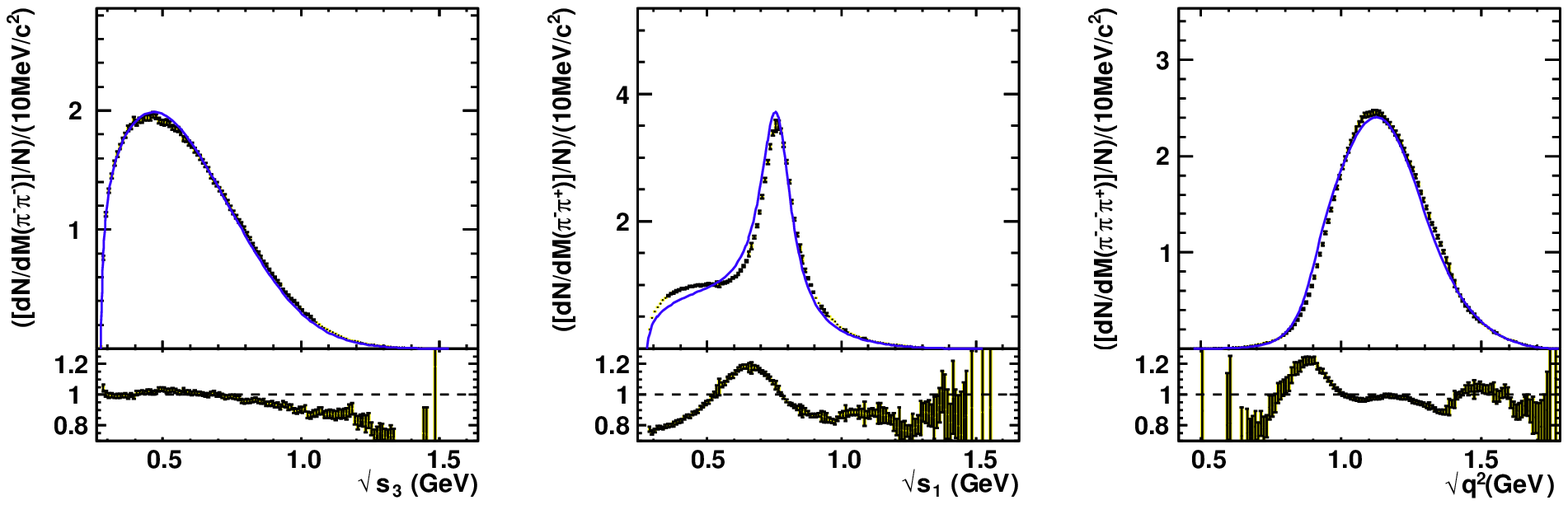}
\includegraphics[width = 1.\textwidth]{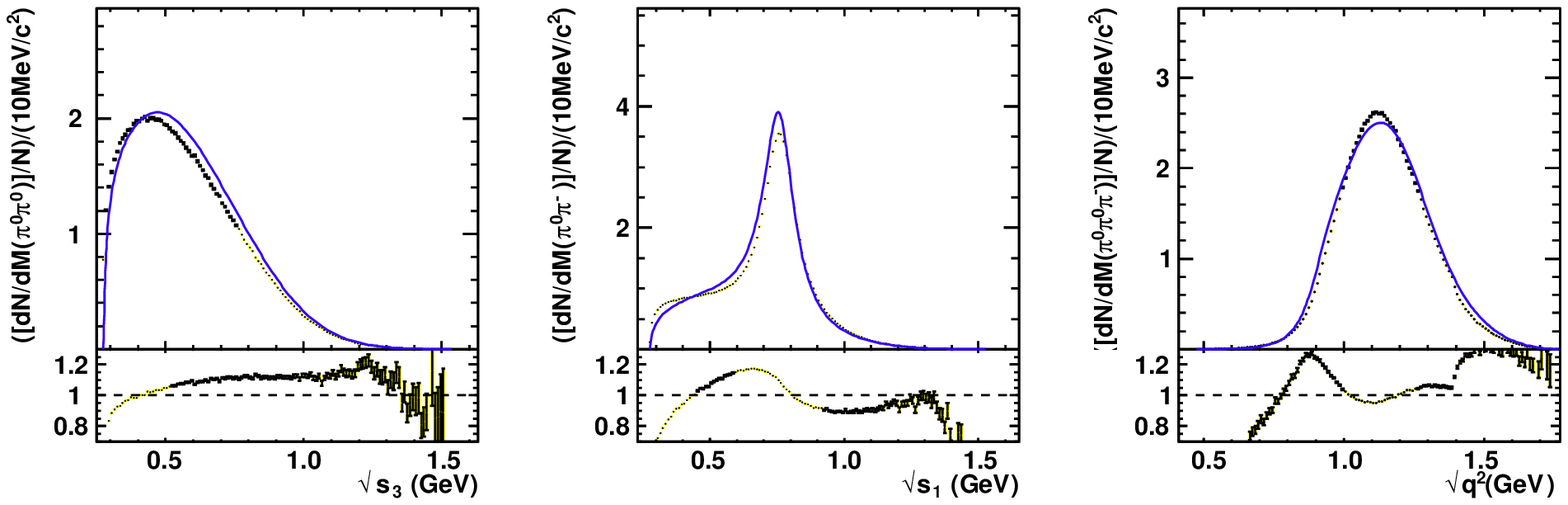}
\end{center}
\caption{{\small Top:
    comparison between the BaBar data
    and our theoretical
    prediction for the $\pi^-\pi^-\pi^+$ decay mode.
    Bottom: comparison between the CLEO 'emulated' data
    (for details the text)
    and our prediction for the $\pi^0\pi^0\pi^-$ decay mode.
    }}
\label{fig.tauola_theory}
\end{figure}

In addition we have compared the two- and three-meson
invariant mass distributions for our theoretical result
and the experimental data.
For the $\pi^-\pi^-\pi^+$ channel,
we used preliminary BaBar data~\cite{Nugent:2013ij}
(Fig.~\ref{fig.tauola_theory}, top panels).
Due to our lack of access to  the $\pi^0\pi^0\pi^-$ data, they have been 'emulated'
on the basis of the results in Ref.~\cite{CLEO:1999}:
Tauola was run with CLEO's AFF from App.~A.1 of~\cite{CLEO:1999} and nominal
fit parameters specified therein in Table III.~\footnote{
We thank J.~Zaremba for providing the corresponding  unnormalized CLEO distributions.  }
The comparison of our parametrization to this
  `emulation' of
CLEO data is shown in  Fig.~\ref{fig.tauola_theory}, bottom panel.

To produce the theoretical distributions the tensor and scalar resonance parameters were fixed
to their value specified in Secs.~\ref{sec:scalar-width} and~\ref{sec:tensor-width}
whereas the vector and axial-vector parameters were fixed to their fit values
in~\cite{Nugent:2013hxa}. All parameters are
summarized in Table~\ref{tab:num_value}
except $c_m$. This coupling,
whose effects are suppressed by $m_\pi^2$ factors, is extracted from the $c_d$ and $F_\pi$ values in Table~\ref{tab:num_value} and the
short-distance constraint $4c_d c_m=F_\pi^2$~\cite{Jamin:2001zq}.

\begin{table}[h!]
\caption{Numerical values of the parameters used to produce the theoretical spectra
in~\ref{fig.tauola_theory}.
All the parameters are in GeV units except for $c_{\sigma}$ and $c_{f_0}$,
which are dimensionless.
}
\label{tab:num_value}
\begin{center}
\begin{tabularx}{\textwidth}{|X|X|X|X|X|X|X|X|}
\hline
$M_{\rho}$  & $M_{\rho'}$   & $\Gamma_{\rho'}$ &  $M_{a_1}$  & $M_{\sigma}$  &  $M_{f_2}$ & $\Gamma_{f_2}$ &$F_{\pi}$
\\
\hline
$ 0.772 $  & $ 1.35 $   & $ 0.448 $ &  $ 1.10 $  & $ 0.8064 $  &  $ 1.275 $ & $ 0.185 $ & $0.0922 $
\\
\hline
\end{tabularx}
\\
\begin{tabularx}{\textwidth}{|X|X|X|X|X|X|X|X|}
\hline
 $F_{V}$   & $F_{A}$ &  $\beta_{\rho}$  & $g_{T}$  &  $c_d$ & $c_{\sigma}$ & $M_{f_0}$ & $c_{f_0}$
\\
\hline
 $ 0.168 $   & $ 0.131 $ &  $ - 0.32 $  & $ 0.028 $  &  $ 0.026 $ & $ 76.12 $ & $1.024$ & $17.7$
\\
\hline
\end{tabularx}
\end{center}
\end{table}

These plots in Fig.~\ref{fig.tauola_theory} are an illustration of our model,
which demonstrates that, even without fitting,
the model qualitatively reproduces the experimental
spectra. No large unwanted deviation from data occurs,
being these values an appropriate
starting point for a more detailed study.
The tuning of our model parameters and the fitting to the data
will be done in a future work~\cite{new-paper-fit}.

\begin{figure}[!t]
\begin{center}
\includegraphics[width = 0.31\textwidth,clip]{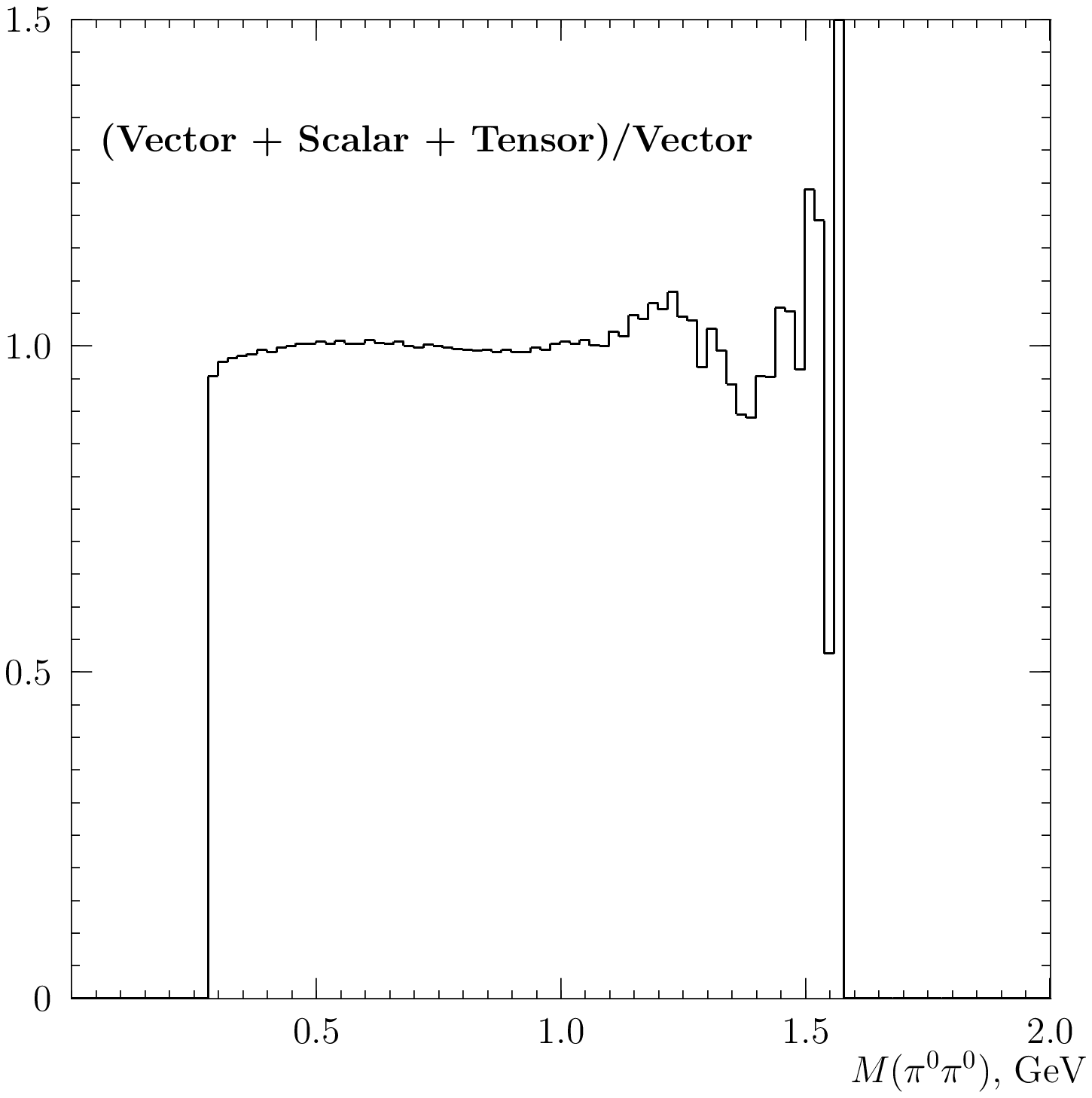}
\hspace*{0.2cm}
\includegraphics[width = 0.32\textwidth,clip]{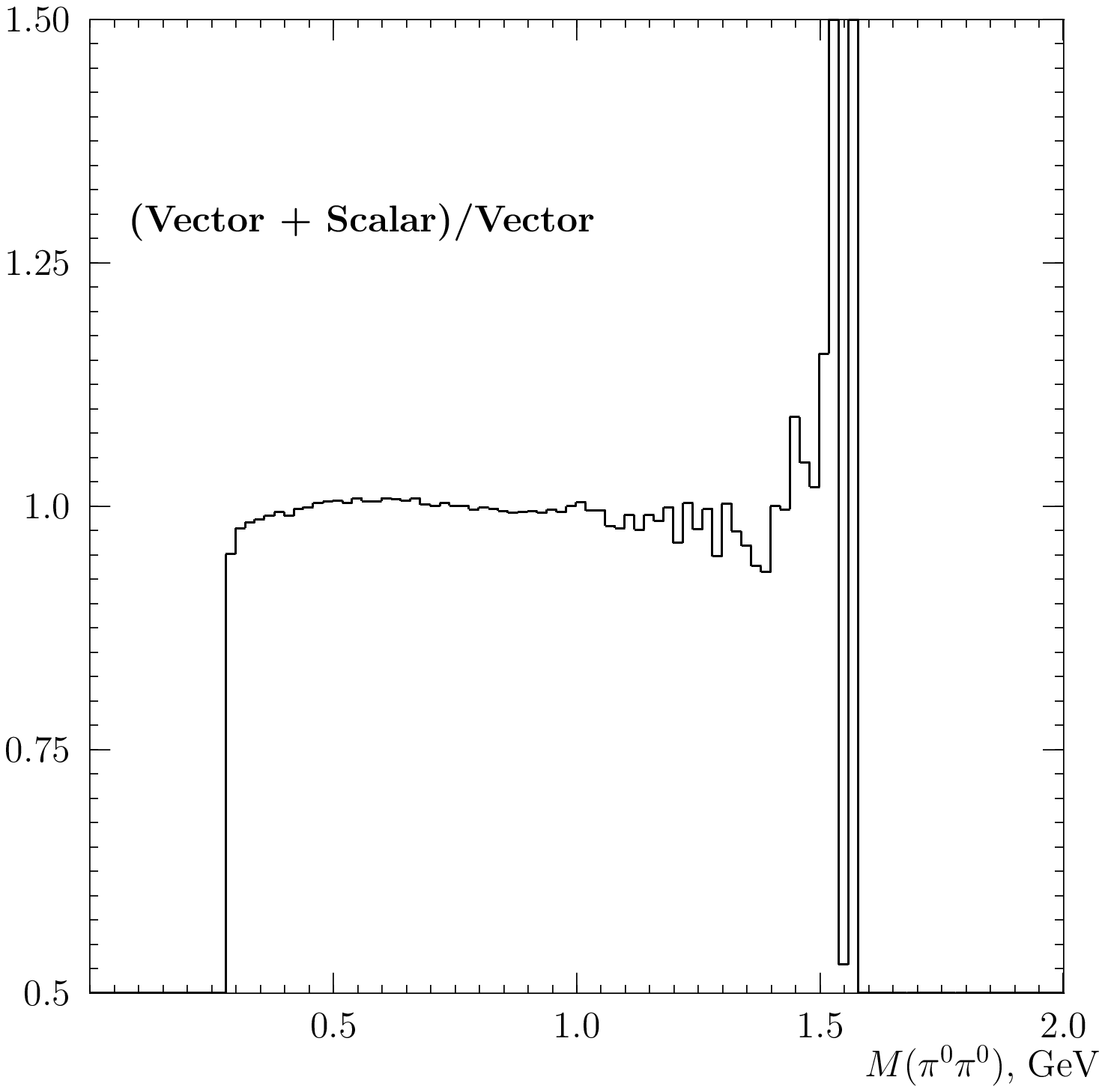}
\hspace*{0.2cm}
\includegraphics[width = 0.31\textwidth,clip]{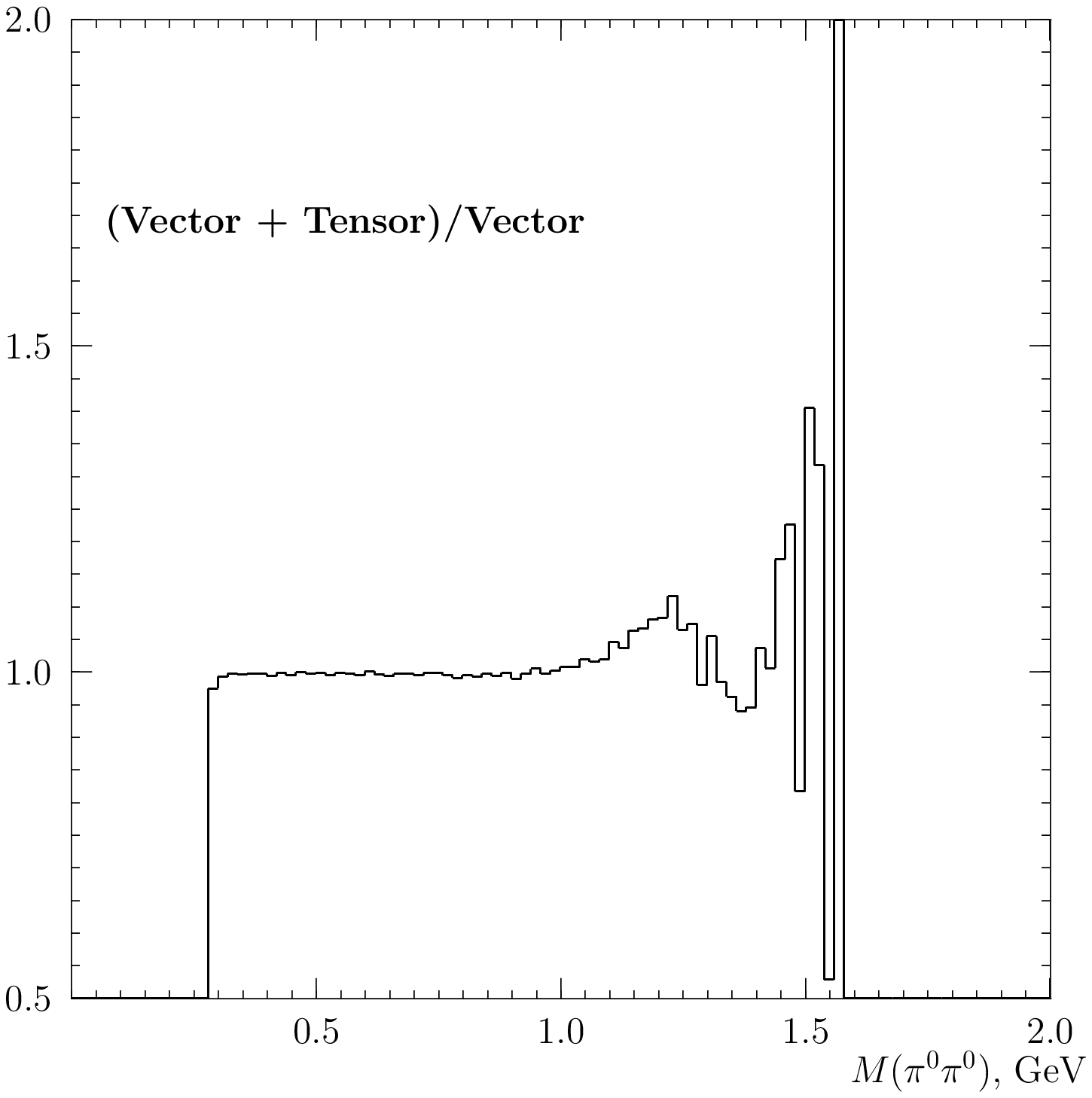}
\end{center}
\caption{{\small
a) Ratio of the vector+tensor+scalar and only vector
$\sqrt{s_3}=M_{\pi^0\pi^0}$ spectral function for $\tau\to\nu_\tau \pi^0\pi^0\pi^-$;
b) Ratio of vector+scalar and only vector;
c) Ratio of vector+tensor and only vector.
All the plots use the inputs in Table~\ref{tab:num_value} ($c_d=26$~MeV).
    }}
\label{fig:VST-vs-V}
\end{figure}

In order to understand the impact of the different contributions we
focus our attention in the $\pi^0\pi^0\pi^-$ channel, where the various contributions
are more neatly separated: vectors only resonate in the $s_1$ and $s_2$ spectra,
and scalars and tensors only resonate in the $s_3$ distribution.
The first thing to notice is that all the distributions are dominated
by the vector contribution ``$V$''
(Lagrangian with only chiral Goldstones,
vectors and axial-vectors~\cite{Dumm:2009va,Shekhovtsova:2012ra}).
The scalar resonances (in particular the $\sigma$ meson)
serve to cure the discrepancies with respect to the data that appear
in the low energy regions, $M_{\pi\pi}< M_\rho$~\cite{Nugent:2013hxa}.
In Fig.~\ref{fig:VST-vs-V} we show the ratio of our
theoretical $\sqrt{s_3}$ distribution including
only the vector contribution $V$~\cite{Nugent:2013hxa}) and
its full result ($V+S+T$) in Fig.~\ref{fig.tauola_theory}
(all with the inputs given in Table~\ref{tab:num_value}).
For this set of parameters, we find that the scalar corrections
are smaller than 10\% in the low-energy region. Therefore,
when fitting the experimental data in this range, we will find that small
variations in the vector parameters may compensate
large modifications in the scalar ones,
being highly correlated for this observable.
Finally, the tensor resonance produces in general a negligible effect
in all the distributions except in the $\sqrt{s_3}$ one around 1.25~GeV,
where one can observe the clear emergence of the $f_2(1270)$ structure
in Fig.~\ref{fig:VST-vs-V}.
\begin{figure}[!t]
\begin{center}
\includegraphics[width = 0.35\textwidth,clip]{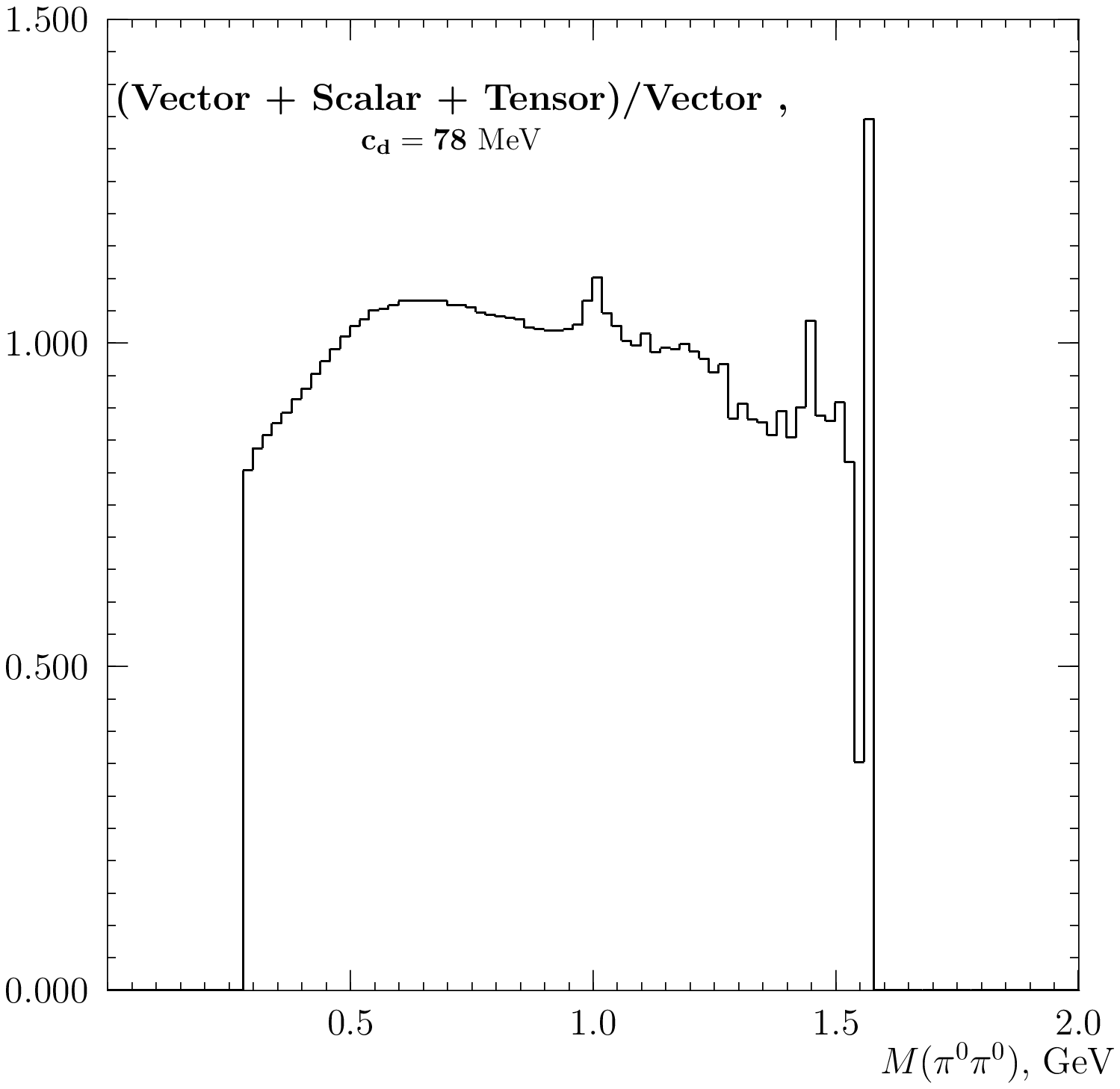}
\hspace*{0.9cm}
\includegraphics[width = 0.34\textwidth,clip]{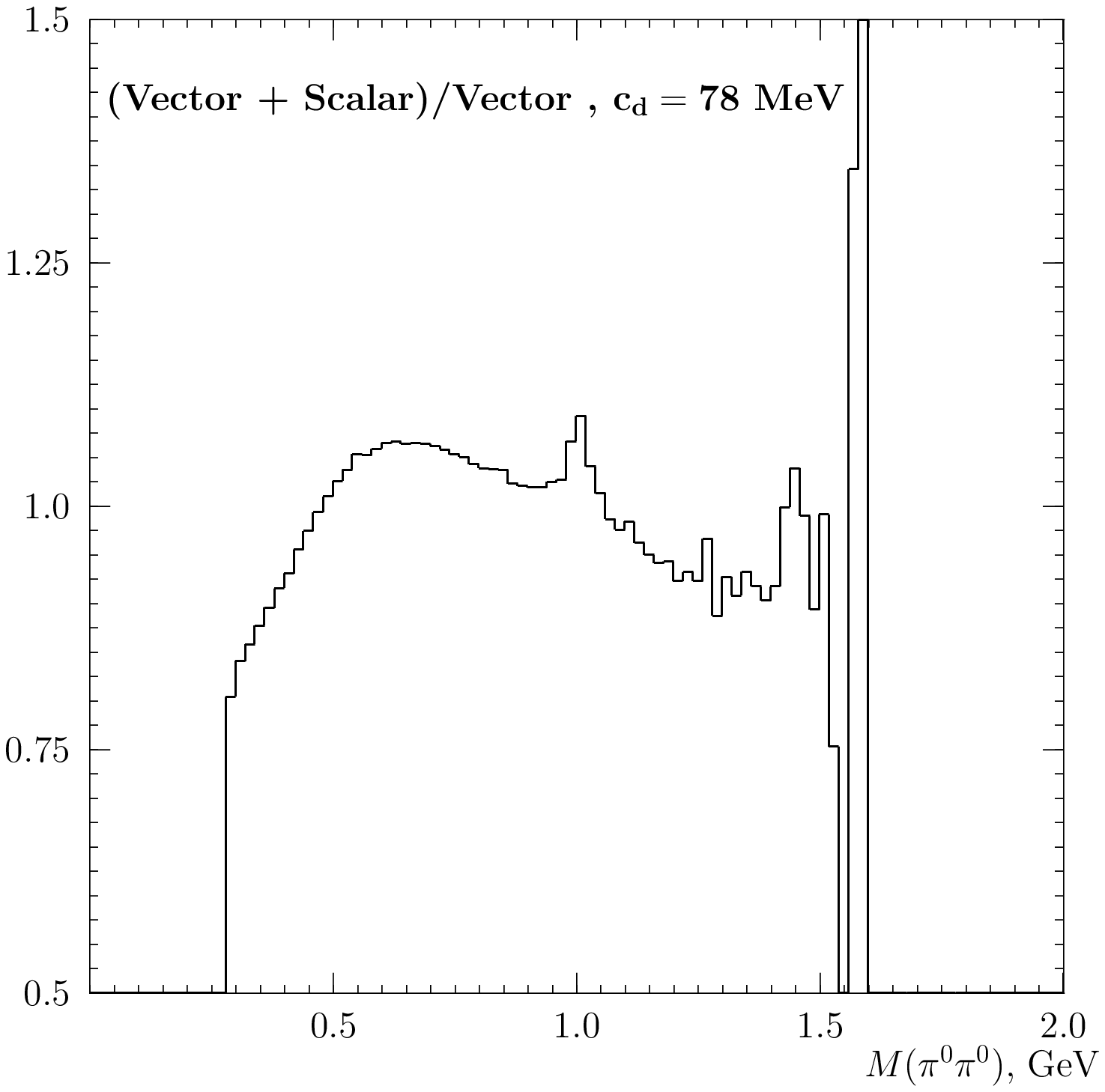}
\end{center}
\caption{{\small
a) Ratio of the vector+tensor+scalar and only vector
$\sqrt{s_3}=M_{\pi^0\pi^0}$ spectral function for $\tau\to\nu_\tau \pi^0\pi^0\pi^-$;
b) Ratio of vector+scalar and only vector.
All the plots use the inputs in Table~\ref{tab:num_value}
except for $c_d$, which is set to $78$~MeV.
The ratio vector+tensor/vector is independent of $c_d$
and is provided in Fig.~\ref{fig:VST-vs-V}.c.
    }}
\label{fig:VST-vs-V-cd=78}
\end{figure}

All the former analyses in this article are performed for the $S\pi\pi$ coupling
$c_d=26$~MeV in Table~\ref{tab:num_value}.
It is not clear whether this is the most suitable value, as other studies
do not lead to a conclusive estimate, allowing a much higher coupling~\cite{Sdecays}.
Since the scalar contribution to the amplitude is essentially proportional to $c_d^2$,
multiplying the value of $c_d$ by a factor 3 increases the impact of the scalar
in the spectral function by one order of magnitude
(through the interference with the $V$ contribution).
For illustration, in Fig.~\ref{fig:VST-vs-V-cd=78}, we show the same ratio
as in Fig.~\ref{fig:VST-vs-V} but for $c_d=78$~MeV.
The impact of these variations can be as important as small modifications
of the $V$ parameters.
Thus, it is not possible to pin down the scalar couplings without an accurate determination
of the vector ones. A joint fit is mandatory.

\begin{figure}[!t]
\begin{center}
\includegraphics[width = 0.45\textwidth]{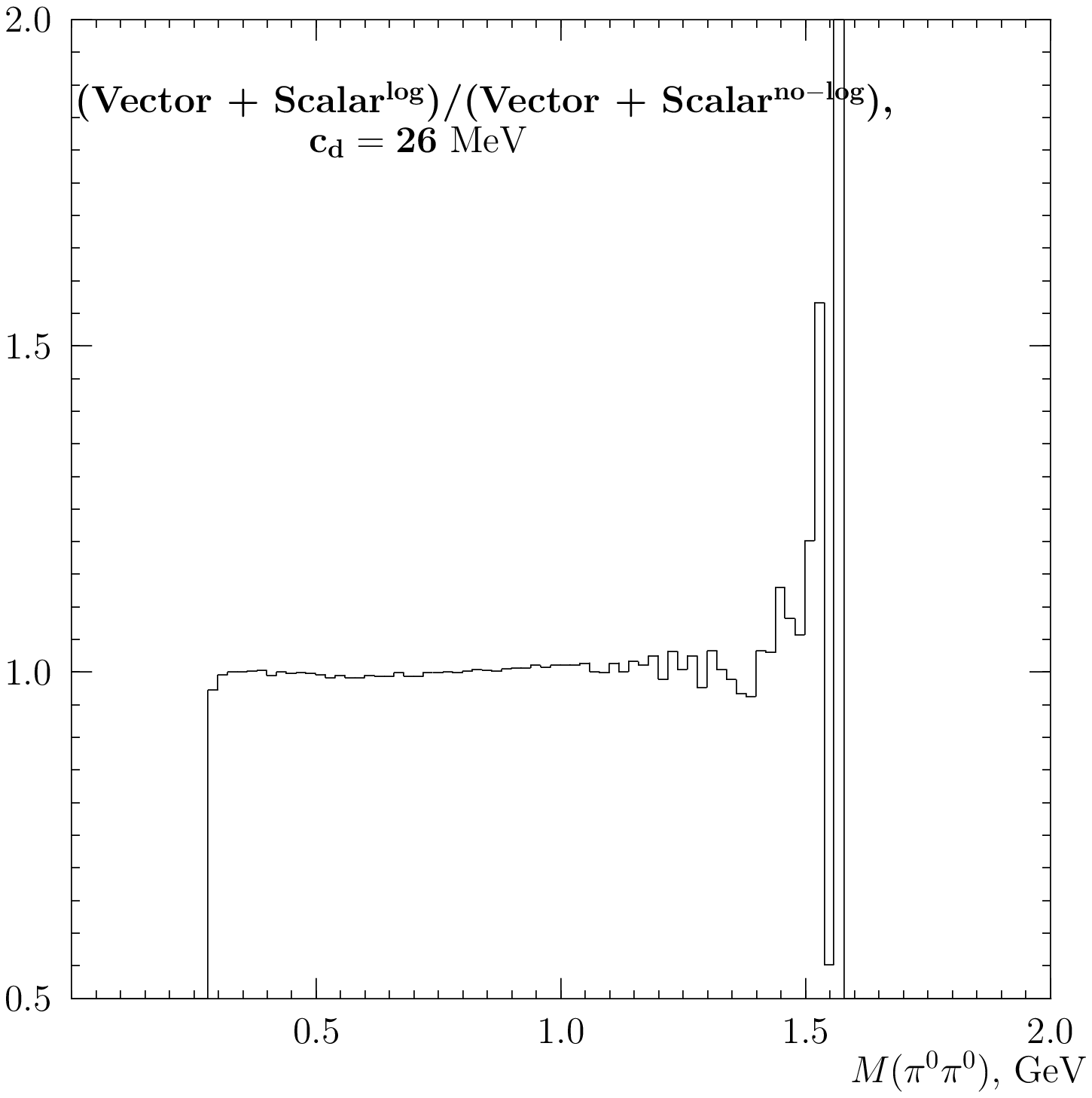}
\hspace*{0.2cm}
\includegraphics[width = 0.45\textwidth]{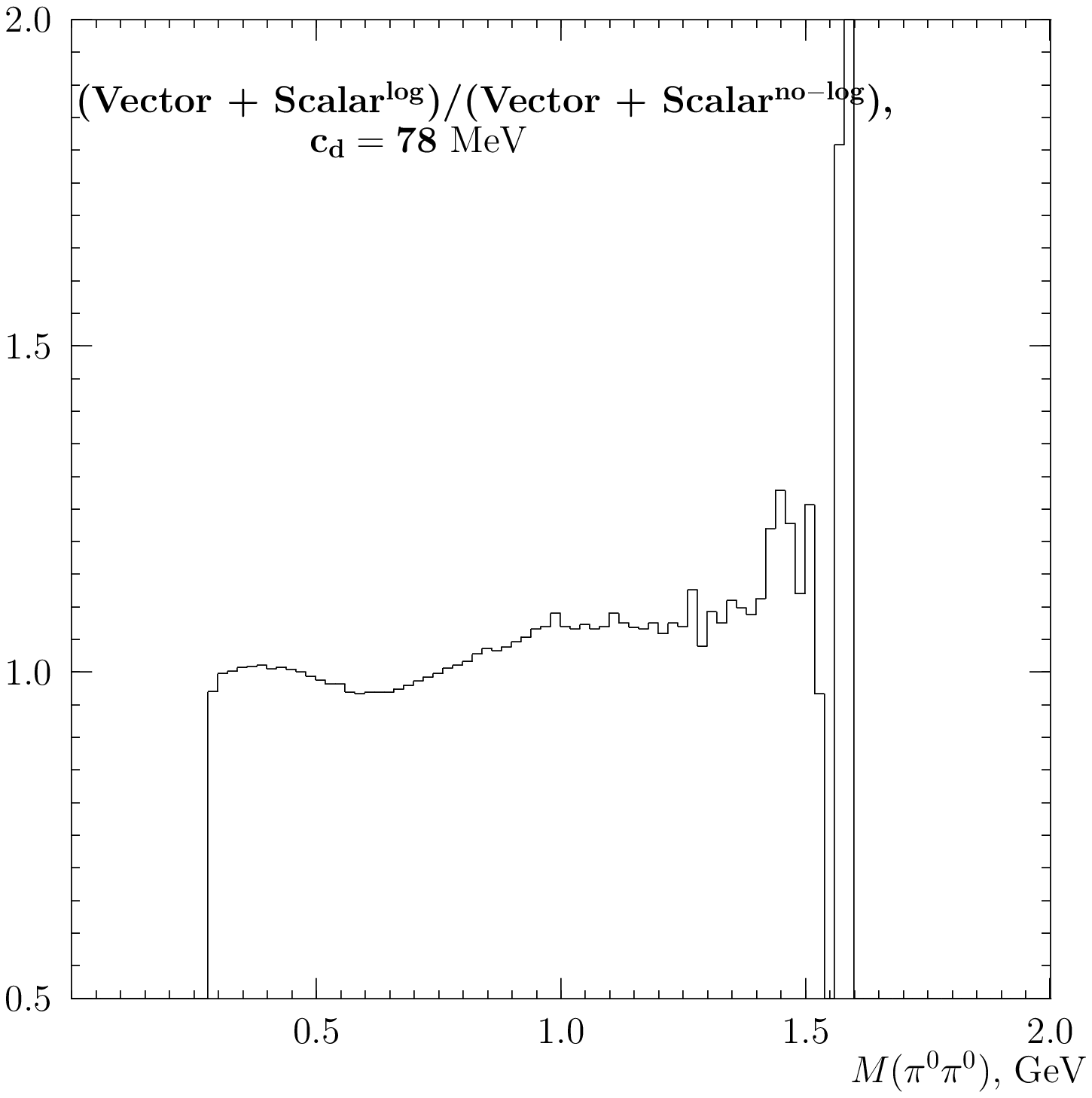}
\end{center}
\caption{{\small
Plots for the ratios of the $\sqrt{s_3}=M_{\pi^0\pi^0}$ spectral functions
for $\tau\to\nu_\tau \pi^0\pi^0\pi^-$:
a) ratio of the full result and the spectral function without the real part of the logs in
the $\sigma$ propagator for $c_d=26$~MeV;
b) ratio of the full result and the spectral function without the real part of the logs in
the $\sigma$ propagator for $c_d=78$~MeV.
In order to better pin down the impact of the scalar propagator structure we only consider the vector+scalar
contribution, dropping the tensors.
    }}
\label{fig:GS-vs-BW}
\end{figure}

Another important numerical issue refers to the relevance
of the real part of the logarithm that is incorporated to the
$\sigma$ propagator \`a la Gounaris-Sakurai.
In Fig.~\ref{fig:GS-vs-BW}.a (Fig.~\ref{fig:GS-vs-BW}.b)
we show the ratio of our
theoretical $\sqrt{s_3}$ distribution neglecting the real part
of the $\sigma$ logs in Eqs.~(\ref{eq.S-propagator2})--(\ref{eq.S-self-energy})
and the full results from these equations
for $c_d=26$~MeV ($c_d=78$~MeV).
For all the other parameters we use the inputs from Table~\ref{tab:num_value}
and take only the vector+scalar contributions for sake of clarity.
Since the scalar contribution is quite small, the impact of the real logs
of the $\sigma$ propagator in the full spectral distributions
is quite suppressed for this $\tau$ decay.
%
%
We want to emphasize that although a Breit-Wigner $\sigma$ can provide an equally good
description of the data~\cite{Nugent:2013hxa},
the aim of the present analysis of the $\sigma$ \`a la Gounaris-Sakurai
is rather to improve the theoretical understanding of broad resonances
within a Lagrangian formalism and its matching to $\chi$PT at low energies.

\section{Conclusions}
\label{sec:conclusions}

In this article we have computed the contribution of scalar and tensor resonances
to the $\tau\to \pi\pi\pi\nu_\tau$ decay axial-vector form-factors.
We have made use of a chiral invariant Lagrangian including
the relevant axial-vector, scalar and tensor resonances
together with the chiral (pseudo) Goldstones.

As a consequence of this,
the chiral symmetry is automatically incorporated  in our
result. This ensures
the proper low-energy matching with $\chi$PT
and that
the currents for $\pi^0\pi^0\pi^-$ and $\pi^-\pi^-\pi^+$ channels
are related as prescribed by isospin symmetry~\cite{Girlanda:1999fu,Finkemeier:1996}.
In addition,
the tensor resonance contribution to the axial-vector current is transverse
in the chiral limit,
improving previous descriptions~\cite{Castro:2011zd}.
A similar thing applies to the scalar contributions.
Chiral symmetry also guaranties the proper low-energy matching with $\chi$PT,
fixing some issues in former parametrizations~\cite{CLEO:1999} (see App.~\ref{app.cleo}).

In addition, the tensor and scalar resonance contributions
to the tau decay are further refined by demanding the
appropriate asymptotic high-energy QCD behaviour for meson form-factors
prescribed by the quark-counting rules~\cite{Brodsky:1973kr}.
 As described in Secs.~\ref{sec:3pi-from-S},~\ref{sec:AFF}
and App.~\ref{app.optical-theorem},
these large--$N_C$ short distance conditions constrain the resonance parameters
of the $T\pi$ and $S\pi$ AFFs, which are essentially determined in terms of the $g_T$
and $c_d$ couplings, respectively, and the resonance masses.

We have also studied an alternative approach
to the sigma description incorporating an analytical description of the width
\`a la Gounaris-Sakurai~\cite{GS-rho}:
instead of just the imaginary part $i\rho_\pi(s)$
required by unitarity in the K-matrix formalism or the Breit-Wigner form~\cite{Nugent:2013hxa},
we considered the full logarithm from the analytical Chew-Mandelstam
dispersive integral~\cite{Chew:1960iv}
or the renormalized two-propagator Feynman integral $\overline{B}_0$.
This parametrization of the $\sigma$ propagator provided a successful description
of the $\eta’\to\eta\pi\pi$ data and its $s$--wave $\pi\pi$ rescattering~\cite{Escribano:2010wt}.
Although it requires further refinements,
we find the exploration of this approach for $\tau\to 3\pi \nu_\tau$ worthy,
as it may help to understand whether it is possible or not to use
a Lagrangian formalism based on a perturbative expansion ($1/N_C$ in our case)
for the description of broad resonances.

We would like to note that in this article
we have considered for the first time the axial-vector--tensor interaction within the
Resonance Chiral Theory approach, extending the work
of Ecker and Zauner on tensors~\cite{Zauner:2007}. We plan to
include vector--tensor interactions in a similar way
in a future paper~\cite{new-paper}
dedicated to the study of the $e^+e^-\to a_2\pi$ process.

We have compared our outcome for the $\pi\pi\pi$ AFF with
former parametrizations with CLEO~\cite{CLEO:1999} and Castro-Mu\~noz~\cite{Castro:2011zd}.
While we coincide on the resonance region, our result incorporates an appropriate
low and high-energy behaviour, improving these works in the latter regimes.
As we plan to incorporate these new results in the Tauola generator, which generates
events from the three pion threshold up to roughly the tau mass, it is important to handle as best as possible the
various energy ranges (low, resonant and high). Some first simulations with the
Tauola Monte Carlo have been provided in Sec.~\ref{sec:Tauola}.
This article is only a preliminary illustration of our resonance chiral Lagrangian approach.
A more thorough numerical analysis is postponed for a future work~\cite{new-paper-fit}.
In order to obtain a good fit to the BaBar data, we will probably need not only
the one-dimensional distributions but also the Dalitz plot. A proper tuning
of the Monte Carlo parameters (e.g., the $S\pi\pi$ coupling $c_d$) should be reading before
the beginning of the Belle-II data taking.

To conclude:  we
would like to remind that the forthcoming project Belle-II~\cite{Abe:2010gxa}
has a broad program devoted to  $\tau$-physics. By 2022, they expect to
record a $50$ times lager data sample than the Belle experiment.
It will give us an opportunity to measure both $\pi^-\pi^-\pi^+$ and $\pi^0\pi^0\pi^-$ decays
and study their intermediate
production mechanisms like, e.g., the tiny contribution from the $f_2\pi^-$ channel. This
will allow us to test our hadronic model and
the isospin symmetry relation between $\pi^-\pi^-\pi^+$ and $\pi^0\pi^0\pi^-$ form factors.

\section*{Acknowledgements}

We are thankful to G. Ecker, G. L\'opez-Castro and P.~Roig
for their helpful comments and feedback on the draft.
We thank J.~Zaremba for providing us the CLEO 'emulated' spectra
and R.~Escribano and T.~Przedzinski for useful discussions.
This work was partly supported
 by the Spanish MINECO fund FPA2016-75654-C2-1-P.

\appendix

\section{Axial-vector form-factor into $3\pi$ in $\chi$PT}
\label{app.ChPT}

In this Appendix, we will focus on the non-chirally suppressed form-factor $\mF_1$.
At tree-level, $\chi$PT gives the low-energy expansion up to $\cO(p^4)$~\cite{Finkemeier:1996}~\footnote{
The relations between $SU(2)$ and $SU(3)$ chiral couplings (respectively, $\bar{\ell}_i$ and $L_i$)
can be found in Sec.~11 of Ref.~\cite{Gasser:1984gg}.  }
\bear
\mF_1^{\pi^0\pi^0\pi^-}(s_1,s_2,q^2)  &=&
 \Frac{2
\sqrt{2}}{3F_{\pi}}
\bigg( 1\,+\, \Frac{4 (2L_1+L_3) }{F_{\pi}^2}(s_3-2m_\pi^2)
\nn\\
&&
\quad + \Frac{4L_2}{F_{\pi}^2} (s_2-2s_1+ 2 m_\pi^2) + \Frac{4 (2L_4+L_5) m_\pi^2}{F_{\pi}^2}
+\Frac{ 2 L_9 q^2}{F_{\pi}^2}\bigg)
\, ,
\label{eq.chpt-AFF}
\eear
with $q^2=(p_1+p_2+p_3)^2$ and $s_1=(p_2+p_3)^2$, etc.
Notice the kinematical constraint $s_1+s_2+s_3=q^2+3m_\pi^2$.

At $\cO(p^2)$ the $\pi^0\pi^0\pi^-$ and $\pi^-\pi^-\pi^+$ channels are
related through isospin in the simple form
\bear
\mF_1^{\pi^0\pi^0\pi^-}(s_1,s_2,q^2)  \,=\,
-\, \mF_1^{\pi^-\pi^-\pi^+}(s_1,s_2,q^2)   &=&
\Frac{2   
\sqrt{2}}{3F} \, .
\eear
Nonetheless, resonance contributions will show up at $\cO(p^4)$
and higher~\cite{rcht,op6rxt,Zauner:2007}, in general spoiling this relation.

In the case when there are only vector contributions to the LECs one finds~\cite{rcht},
\bear
 L_2\bigg|_V=2 L_1\bigg|_V= -\Frac{L_3}{3}\bigg|_V=\Frac{G_V^2}{4 M_V^2}\, , \qquad
 L_9\bigg|_V= \Frac{F_V G_V}{2 M_V^2}\, ,
\eear
with the remaining $\cO(p^4)$ LECs being zero.
Thus, one has the $\cO(p^4)$ contribution~\cite{Zuo:2013}
\bear
\mF_1^{\pi^0\pi^0\pi^-}(s_1,s_2,q^2) \bigg|_V  \,=\,
-\, \mF_1^{\pi^-\pi^-\pi^+}(s_1,s_2,q^2)  \bigg|_V  &=&
\Frac{2
\sqrt{2}}{3F}
\left(
\Frac{8 L_1 (3 s_2 -2 q^2)}{F^2}
+\Frac{ 2 L_9  q^2}{F^2}\right)\bigg|_V
\, .
\nn\\
\eear

The situation is different in the case when there are only scalar contributions to the $\cO(p^4)$
LECs~\cite{rcht}:
\bear
L_1\bigg|_S= \Frac{\widetilde{c}_d^2}{2 M_{S_1}^2} -  \Frac{c_d^2}{6 M_S^2}\, , \qquad
L_2\bigg|_S
= L_9\bigg|_S =  0\, , \qquad
L_3\bigg|_S=  \Frac{c_d^2}{2 M_S^2}\, ,
\nn\\
L_4\bigg|_S= \Frac{\widetilde{c}_d \widetilde{c}_m}{M_{S_1}^2} -  \Frac{c_d c_m}{3 M_S^2}\, ,\qquad
L_5\bigg|_S= \Frac{c_d c_m}{M_S^2}\, .
\eear
Taking this into account one obtains the $\cO(p^4)$ contribution
\bear\label{eq:scal-contr-limit}
\mF_1^{\pi^0\pi^0\pi^-}(s_1,s_2,q^2) \bigg|_S&=&
 \Frac{2 \sqrt{2}}{3F_{\pi}}
\bigg(
\Frac{4 (2L_1+L_3) }{F_{\pi}^2}(s_3-2m_\pi^2)  + \Frac{4 (2L_4+L_5) m_\pi^2}{F_{\pi}^2} \bigg)\bigg|_S
\, ,
\nn\\
\mF_1^{\pi^-\pi^-\pi^+}(s_1,s_2,q^2)\bigg|_S  &=&
 - \Frac{2 \sqrt{2}}{3F_{\pi}}
 \bigg(
\Frac{4 (2L_1+L_3) }{F_{\pi}^2}(2 s_1 - s_2 -2 m_\pi^2)
+ \Frac{4 (2L_4+L_5) m_\pi^2}{F_{\pi}^2} \bigg)\bigg|_S\, .
\nn\\
\eear
Except for the special point $s_1=(q^2+3 m_\pi^2)/3$, the $\mF_1$ functions of the two decay channels
have a different kinematical dependence and one cannot simply assume
$\mF_1^{\pi^0\pi^0\pi^-}(s_1,s_2,q^2) \,=\,
-\, \mF_1^{\pi^-\pi^-\pi^+}(s_1,s_2,q^2)$.
This precise expression~(\ref{eq:scal-contr-limit})
can be directly obtained from the low-energy limit of Eq.~(\ref{eq.ff-scal}),
\bear
\mF_1^{\pi^0\pi^0\pi^-}(s_1,s_2,q^2) \bigg|_S &=&
\frac{4\sqrt{2}}{3 F_\pi^3}\frac{1}{M_S^2}\, [ c_d^2 (s_3 - 2m_\pi^2)
 + 2 c_d c_m m_\pi^2]  \, ,
\eear
where in the large $N_C$ limit the octet and singlet scalar couplings are related in the
form $\widetilde{c}_d=c_d/\sqrt{3}$ and $\widetilde{c}_m=  c_m/\sqrt{3}$,
and $L_1|_S$ and $L_4|_S$ turn zero~\cite{rcht}.

Taking only the tensor resonance contribution,
the $\cO(p^4)$ contributions to the form-factors  become
\bear\label{eq:tens-contr-limit}
\mF_1^{\pi^0\pi^0\pi^-}(s_1,s_2,q^2) \bigg|_T  &=&
 \Frac{2 \sqrt{2}}{3F_{\pi}}
\bigg(
\Frac{4 L_3 }{F_{\pi}^2}(s_3-2m_\pi^2) \bigg)\bigg|_T \, ,
\nn\\
\mF_1^{\pi^-\pi^-\pi^+}(s_1,s_2,q^2) \bigg|_T  &=&
 - \Frac{2   \sqrt{2}}{3F_{\pi}}
\bigg(
\Frac{4 L_3 }{F_{\pi}^2}(2 s_1 - s_2 -2 m_\pi^2) \bigg)\bigg|_T\, ,
\eear
with the $\cO(p^4)$ chiral low-energy constants~\cite{Zauner:2007},
\bear
L_1\bigg|_T =   L_2\bigg|_T =  0 \, ,
\qquad\qquad
L_3\bigg|_T  = \Frac{ g_T^2}{3 M_T^2}\, ,
\label{eq:Ecker-Zauner-rel}
\eear
and zero for all the remaining LECs.
As it happened in the scalar resonance case,
the relation
 $\mF_1^{\pi^0\pi^0\pi^-}(s_1,s_2,q^2) \,=\,
-\, \mF_1^{\pi^-\pi^-\pi^+}(s_1,s_2,q^2)$
is generally not true, only being fulfilled at the special kinematical point
$s_1=(q^2+3 m_\pi^2)/3$.
The result~(\ref{eq:tens-contr-limit})
can be obtained directly from the determination~(\ref{eq.ff-tens}): the
$\cO(p^4)$ term in the low-energy expansion of our tensor-exchange prediction
is given by the diagrams $a$ and $b$ in Fig.~\ref{fig.diagr}
(with their subsequent $T\to\pi\pi$ decay),
\bear
\mF_{1,\,\, (RSD)}^{\pi^0\pi^0\pi^-}(s_1,s_2,q^2)
&=&
 \Frac{8\sqrt{2} g_T^2}{9 F_\pi^3 M_T^2}
\left(-6 s_1 +3 s_2 -2s_3 +10 m_\pi^2\right)\,  ,
\eear
and those in Fig.~\ref{fig.diagr-nonR},
\bear
\mF_{1,\,\, (0)}^{\pi^0\pi^0\pi^-}(s_1,s_2,q^2) &=&
 \Frac{8\sqrt{2} g_T^2}{3 F_\pi^3 M_T^2} \left(2 s_1 -s_2 +s_3-4 m_\pi^2\right) \, .
\eear
The remaining contributions to $\mF_1(s_1,s_2,q^2)$ are zero at $\cO(p^4)$.
Therefore the total contribution at that chiral order is
\bear
\mF_1^{\pi^0\pi^0\pi^-}(s_1,s_2,q^2) \bigg|_T&=&  \Frac{8\sqrt{2} g_T^2}{9 F_{\pi}^3 M_T^2}
\left(s_3-2 m_\pi^2\right) \, .
\label{eq:low-energy-T}
\eear
Matching the expression~(\ref{eq:low-energy-T}) and Eq.~(\ref{eq.chpt-AFF}) one recovers for
$L_{1,2,3}\bigg|_T $ the relations~(\ref{eq:Ecker-Zauner-rel}) from Ref.~\cite{Zauner:2007}.

\section{Optical theorem and axial-vector form-factors}
\label{app.optical-theorem}

The correlator of two axial-vector currents $J_A^{\alpha}=\bar{d} \gamma^\alpha\gamma^5 u$ ,
\bear
\Pi_{AA}(q)^{\mu\nu} &\equiv & i\,\Int {\rm d^4 x} \, e^{i q x} \bra 0| T\{ \, J_A^\mu (x) \, J_A^{\nu}(0)^\dagger  \, \} |0\ket\, ,
\eear
is described by two Lorentz scalar functions, the transverse and longitudinal correlators,  $\Pi_T(q^2)$ and $\Pi_L(q^2)$, respectively:
\bear
\Pi_{AA}(q)^{\mu\nu}  &=& - q^2 P_T(q)^{\mu\nu}  \, \Pi_T(q^2)\,\,\,
+\,\,\, q^2 P_L(q)^{\mu\nu} \Pi_L(q^2)\, .
\eear
The conservation of the axial-vector current in the chiral limit implies that $\Pi_{L}(q^2)$
is suppressed by the up and down quark mass combination $(m_u+m_d)$, this is, by $m_\pi^2$.

The axial-vector form-factors for the production of a generic state $X$ and its corresponding hadronic matrix element,
\bear
H^\alpha =\bra X |\,\bar{d}\gamma^\alpha\gamma_5 u\, |0\ket \,,
\eear
determines the contribution to the spectral functions of $\Pi_{T,L}$ from that absorptive cut
through the optical theorem. For a two-particle intermediate state $X$ with masses $m_1$ and $m_2$ one has
\bear
{\rm Im}\Pi_T(t)\bigg|_{\rm cut\, X} &=& \, - \,
\left(\Frac{\lambda(t,m_1^2,m_2^2)^{\frac{1}{2}} }{48\pi  t^2}\right)
\,\, \sum_{\rm helicities} H_\alpha P_T(q)^{\alpha\beta}  H^*_\beta \, ,
\nn\\
{\rm Im}\Pi_L(t)\bigg|_{\rm cut\, X} &=&
\left(\Frac{\lambda(t,m_1^2,m_2^2)^{\frac{1}{2}} }{16\pi  t^2}\right)
\,\, \sum_{\rm helicities} H_\alpha P_L(q)^{\alpha\beta}  H^*_\beta \, ,
\eear
with $t=q^2$, $P_L(q)^{\alpha\beta} =q^\alpha q^\beta/q^2$,
$\lambda(x,y,z)=x^2+y^2+z^2-2 xy -2 xz -2 yz$ and the summation referring to the helicities of the two-particle intermediate state~$X$.

Perturbative QCD tells that the full spectral function goes to a constant at high energies
and thus, the contribution from each (infinitely many) hadronic intermediate
states vanishes for $q^2\to \infty$~\cite{rcht-FFs}. This agrees with Brodsky-Lepage's quark-counting rules for asymptotic behaviour of hadronic form-factor
in the ultraviolet~\cite{Brodsky:1973kr}.

\subsection{$S\pi$ AFF}

The $S_{I=0} \, \pi^-$ absorptive cut contributes to the axial-vector correlator in the form~\cite{rcht-FFs}
\bear
{\rm Im}\Pi_{T}(t)\bigg|_{S\pi} &=& \Frac{\lambda(t,M_S^2,m_\pi^2)^{\frac{3}{2}} \theta(t-t_{th})}{48\pi t^3}
\,\, |\mF^a_{S\pi}(t)|^2\, ,
\label{eq.Spi-spectral-function}
\\
{\rm Im}\Pi_{L}(t)\bigg|_{S\pi} &=& \Frac{\lambda(t,M_S^2,m_\pi^2)^{\frac{1}{2}} \theta(t-t_{th})}{16\pi t}
\,\, |\mH^a_{S\pi}(t)|^2\, ,
\eear
with $t=q^2$ and $t_{th}=(M_{S_{I=0}}+m_\pi)^2$.
In the chiral limit the phase-space factor turns $\lambda(t,M_S^2,0)^{\frac{3}{2}} / t^3=  (1- M_S^2/t)^3$.

Requiring that the contribution to the transverse spectral function vanishes at infinite momentum transfer implies the (minimal) asymptotic behaviour
\bear
\mF^a_{S\pi}(t)\quad &\stackrel{t\to\infty}{\longrightarrow}& \quad \cO\left( \Frac{1}{t}\right)\, .
\eear

\subsection{$T\pi$ AFF}

The $T\pi^-$ cut contributes to the transverse
 spectral function.
The corresponding expressions are rather lengthy but in the chiral limit they become
\bear
{\rm Im}\Pi_{T}(t)\bigg|_{T\pi} &=&
\Frac{\theta(t-M_T^2)}{192\pi} \, \left(1-\Frac{M_T^2}{t}\right)^3
\,  \left[ \,   \Frac{t}{M_T^2} \, |\mF^a_{T\pi}(t)|^2\, +\,
\Frac{t^2}{6 M_T^4} \, \left|  \widetilde{\mG}^a_{T\pi}(t)  \right|^2  \, \right]\, ,
\eear
with
\bear
\widetilde{\mG}^a_{T\pi}(t) &=&  \Frac{t}{2} \left(1-\Frac{M_T^2}{t}\right)^2 \mG^a_{T\pi}(t)\,
-\, \left(1+\Frac{M_T^2}{t}\right) \mF^a_{T\pi}(t) \, ,
\eear
with the phase-space factor in the chiral limit
$\lambda(t,M_T^2,m_\pi^2)\stackrel{m_\pi\to 0}{\longrightarrow}(1-M_T^2/t)^2$.
For the algebra of Lorentz contractions,
we made use of the completeness relation~\cite{Zauner:2007}
\bear
\sum_{\epsilon}   \epsilon_{\mu\nu} \epsilon^*_{\alpha\beta}
&=& \Frac{1}{2}\left(P_{\mu\alpha} P_{\nu\beta} + P_{\nu\alpha} P_{\mu\beta}\right)
\,-\, \Frac{1}{3} P_{\mu\nu} P_{\alpha\beta} \, , \qquad
\mbox{with } P_{\mu\nu}=g_{\mu\nu} -\Frac{k_\mu k_\nu}{M_{f_2}^2}  \, .
\nn\\
\eear

Requiring that the contribution to the spectral function vanishes at infinite momentum transfer implies the (minimal) asymptotic behaviour
\bear
\mF^a_{T\pi}(t)\quad &\stackrel{t\to\infty}{\longrightarrow}& \quad \cO\left( \Frac{1}{t}\right)\, ,
\nn\\
\widetilde{\mG}^a_{T\pi}(t)\quad &\stackrel{t\to\infty}{\longrightarrow}& \quad \cO\left( \Frac{1}{t^2}\right)\, ,
\eear
which implies
\bear
\mG^a_{T\pi}(t)\quad &\stackrel{t\to\infty}{\longrightarrow}& \quad \cO\left( \Frac{1}{t^2}\right)\, .
\eear

The contribution to the longitudinal spectral function from the $T\pi^-$ cut is given by
\bear
{\rm Im}\Pi_{L}(t)\bigg|_{T\pi} &=&
\Frac{\lambda(t,M_T^2,m_\pi^2)^{\frac{5}{2}}  }{384\pi M_T^4 t}
\, |\mH^a_{T\pi}(t)|^2 \, .
\eear
The longitudinal form-factor $\mH^{a}_{T\pi}$
is chirally suppressed by $m_\pi^2$
and must have a minimal asymptotic fall off,
\bear
\mH^a_{T\pi}(t)\quad &\stackrel{t\to\infty}{\longrightarrow}& \quad \cO\left( \Frac{m_\pi^2}{t^3}\right)\, .
\eear

\section{Comparison with other production analyses}
\label{app.comparisons}

\subsection{Comparison with CLEO~\cite{CLEO:1999}   }
\label{app.cleo}

We now compare our expression for the hadronic current (\ref{eq.Htens}) with the corresponding
theoretical expression used by CLEO for the $f_2$ production
(Eq. (A3) in Ref.~\cite{CLEO:1999}).
In the chiral limit
the latter is
\bear
H^\mu &=& \, -\,  \Frac{ i\, \beta_5 M_T^2}{ (M_A^2 -q^2)\, (M_T^2-s_3)}
\, F_{R_5}
\,P_T(q)^{\mu\nu}\, \bigg[ (q\Delta p)\, \Delta p_\nu
\, + \, \Frac{(\Delta p)^2}{3 s_3}   (qk)\, k_\nu \bigg] \, ,
\nn\\
\label{eq:CLEO}
\eear
where the $a_1$ and $f_2$ widths in the denominators in Ref.~\cite{CLEO:1999}
have been dropped to provide a more transparent comparison with our expressions.
Likewise, we set the axial-vector radius $R_5=0$ and set the momentum dependent function
to the value $F_{R_5}=1$ in the parametrization considered by CLEO
to incorporate finite $a_1$ size effects~\cite{CLEO:1999}.
We have also used $P_T(q)^{\mu\nu} q_\nu=0$
to simplify the expression therein.
Notice that in CLEO's notation $H^\mu=j_5^\mu$.

Our result reproduces that in Ref.~\cite{CLEO:1999} if one keeps just the contribution
$H^\mu_{(2)\, {\rm a_1-pole}}$~(\ref{eq.contri2A}) --with the axial-vector and tensor resonance poles,
respectively in $q^2=M_A^2$ and $s_3=M_T^2$--,
and then sets the high energy condition~(\ref{eq.constraints1}).
Thus, taking just the first two lines of Eq.~(\ref{eq.contri2A})
with the latter condition (the non-singular term with $(M_T^2-s_3)$ is dropped),
one recovers the corresponding expression in Eq.~(A.3) from~\cite{CLEO:1999},
with the identification
$$\beta_5= \Frac{8 g_T F_A \lambda_1^{AT} M_A^2 }{M_T^2 F_\pi^3} .$$

The form-factors  $\mF_1$ and $\mF_P$ derived from Ref.~\cite{CLEO:1999}
can be rewritten as~\footnote{
Note that here we use the form-factor convention
given by Eq.~(\ref{eq.hadr-curr-3pions}).  }
\bear
\mF_1^{\pi^0\pi^0\pi^-}(s_1,s_2,q^2) &=&
\frac{\beta _5 F_{R_5}}{9\left(M_A^2-q^2\right)} \Frac{M_T^2}{\left(M_T^2-s_3\right)}
\left[
5 s_1-4 s_2+s_3  + \Frac{2 m_\pi^2}{s_3} \left(  m_{\pi }^2- q^2  - 2 s_3 \right)
\right] \, ,
\nn\\
\mF_{P}^{\pi^0\pi^0\pi^-}(s_1,s_2,q^2) &=& 0 \, .
\label{eq.CLEO-AFF}
\eear
This expression agrees with our determination in Eq.~(\ref{eq.tens_rsd}):
in our case, after incorporating the high-energy constraints,
one finds that $\mF_1^{\pi^0\pi^0\pi^-}(s_1,s_2,q^2)\approx \mF_{1,RSD}^{\pi^0\pi^0\pi^-}(s_1,s_2,q^2)$
for $s_3 \approx   M_T^2$, showing the structure in (\ref{eq.CLEO-AFF}).

One can see that the parametrization (\ref{eq.CLEO-AFF}) has a subthreshold singularity at $s_3=0$,
absent in the low-energy $\chi$PT  prediction~\cite{Finkemeier:1996} (see App.~\ref{app.ChPT}).
Moreover,
in the chiral limit ($m_\pi\to 0$), the comparison of
 Eqs.~(\ref{eq.CLEO-AFF}) and~(\ref{eq.chpt-AFF})
shows that the coupling $L_9$
must receive a non-zero contribution caused by the tensor resonance.
However, in the chiral limit $L_9$ is the only $\cO(p^4)$ coupling
that appears in the pion vector form-factor at tree-level, \textit{i.e.}
it can never get contributions from spin--2 resonance exchanges.

To conclude: the CLEO parametrization for the tensor resonance contribution to AFF
agrees with the R$\chi$T description
only near the resonance energy region and does not
reproduce the low-energy behaviour predicted by $\chi$PT.

We also compare our results for the scalar contributions to the AFF
with the corresponding CLEO results (Eq.~(3) of~\cite{CLEO:1999}).
Expressing CLEO result in terms of the form-factor convention in~(\ref{eq.hadr-curr-3pions})
one obtains
  \begin{equation}
\mF_1^{\pi^0\pi^0\pi^-}(s_1,s_2,q^2) = \mF_2^{\pi^0\pi^0\pi^-}(s_{\2},s_{1},q^2)
=
\,- \,\Frac{2\beta_S  F_{R_S}   }{3 (M_A^2- q^2)   }
\,\,  \Frac{M_S^2}{M_S^2 - s_3} ,
  \end{equation}
where we have dropped the widths in the denominators for the comparison
and $\beta_S$ is $\beta_6$ or $\beta_7$ depending on whether we refer to $S=\sigma$
or $S=f_0(980)$, respectively.
Likewise, we have set the axial-vector radius $R_S=0$ and set the momentum dependent function
to the value $F_{R_S}=1$ in the parametrization considered by CLEO
to incorporate finite $a_1$ size effects~\cite{CLEO:1999}.
In our case,
after  applying the high-energy constraints,
we got the $S\pi$ form-factor~(\ref{eq.Spi-AFF+HE}) and the three-pion AFF,
\bear
\mF_1^{\pi^0\pi^0\pi^-}(s_1,s_2,q^2) = \mF_2^{\pi^0\pi^0\pi^-}(s_{2},s_{1},q^2)
=
\Frac{4 \sqrt{2}c_d}{3 F_\pi^3}\Frac{M_{A}^2}{M_{A}^2 - q^2}
\Frac{  (  c_d(s_3 - 2m_\pi^2) + 2c_m m_\pi^2  )  }{M_S^2 - s_3 }.
\nn\\
\label{eq.RChT-F1-S}
\eear
This result is later refined by incorporating the $\sigma- f_0(980)$ mixing
through the replacement in~(\ref{eq.MS-splitting}).
Comparing CLEO's expression and ours, we arrive to the conclusion
that the CLEO parametrization for the scalar contribution
to AFF only agrees with the R$\chi$T results
near the scalar resonance region $s_3\approx M_S^2$,
where the numerator of~(\ref{eq.RChT-F1-S}) is approximately constant.

\subsection{Comparison with Castro and Mu\~noz~\cite{Castro:2011zd}}
\label{app.Castro-comparison}

Analysis~\cite{Castro:2011zd} expresses the even intrinsic-parity part of the AFF
into a tensor $T(k,\epsilon)$
and a pseudo-Goldstone $P(p)$ in terms of three independent form-factors
 $\kappa$ and $b_\pm$ (see Eq.~(2) in Ref.~\cite{Castro:2011zd}).
They are related to the form-factors in this work through
\bear
\kappa  =  - i \, \mF^a_{TP}\, ,
\quad
b_+  =  - \Frac{i}{2} \, \mG^a_{TP} \, ,
\quad
b_-  =  \Frac{i}{q^2}\left( \mF^a_{TP} + \left((qp) -\frac{1}{2} q^2\right) \mG^a_{TP}
+ \mH^a_{TP}\right)\, .
\eear

Ref.~\cite{Castro:2011zd} finds $b_+=0$, as in our result in Eq.~(\ref{eq.TP-AFF+constraints}).
In addition, in the chiral limit, requiring these form-factors to fall-off at high energies as
$\kappa(q^2)\stackrel{q^2\to\infty}{\longrightarrow }\cO(1/q^2)$
and
$b_-(q^2)\stackrel{q^2\to\infty}{\longrightarrow }\cO(1/q^4)$,
the relation between the prediction of~\cite{Castro:2011zd}
and ours is given (e.g., for the $f_2\pi^-$ production) by
\bear
\Frac{8 g_T}{F_\pi} \,=\, - \Frac{ f_{a_1} \, g_{f_2 a_1 \pi} }{M_{a_1}} \, ,
\qquad
\Frac{ 8 g_T}{F_\pi} \, =\, -\, F_{\pi} \, g_{f_2\pi\pi}\, .
\eear

\section{Tauola's notation for form factors}
\label{app.relations}
In the Tauola notation the three-pion hadronic current is written:
\bear
\bra 3\pi |\bar{d}\gamma^\mu\gamma_5 u|0\ket &=&  H^{3\pi}(q^2,s_1,s_2)^\mu
\nn\\
&=&  \Frac{i}{F_\pi}  \, P_T^{\mu\nu} (q) \left[
\mF^{\mbox{\tiny Tauola}}_1(q^2,s_1,s_2)\,\, (p_2 - p_3)_\nu \,\,
+\,\, \mF^{\mbox{\tiny Tauola}}_2(q^2,s_1,s_2) \,\,  (p_1-p_3)_\nu   \right]
\nn \\
&& + \,\, \Frac{i \, q_\mu}{F_\pi} \,\,\mF_4^{\mbox{\tiny Tauola}}(q^2,s_1,s_2) \, .
\eear
Therefore, the Tauola form-factors~\cite{Shekhovtsova:2012ra} are related with our
convention in Eq.~(\ref{eq.hadr-curr-3pions}) through
\bear
\mF_1(q^2,s_1,s_2)^{\rm Tauola}\,  &=&\,  F_\pi\, \mF_2(s_1,s_2,q^2) , \nn\\
\mF_2(q^2,s_1,s_2)^{\rm Tauola} &=& \, F_\pi\, \mF_1(s_1,s_2,q^2) , \nn \\
\mF_4(q^2,s_1,s_2)^{\rm Tauola} &=& \, F_\pi \,\mF_P(s_1,s_2,q^2) ,
\eear
with the isospin relations
\bear
\mF^{\mbox{\tiny Tauola}}_1(q^2,s_1,s_2)^{--+}  &=&
\mF^{\mbox{\tiny Tauola}}_1(q^2,s_3,s_2)^{00-} -  \mF^{\mbox{\tiny Tauola}}_1(q^2,s_3,s_1)^{00-}
- \mF^{\mbox{\tiny Tauola}}_1(q^2,s_1,s_3)^{00-} \, ,
\nn
\\
\mF^{\mbox{\tiny Tauola}}_4(q^2,s_1,s_2)^{--+}  &=&
 \mF^{\mbox{\tiny Tauola}}_4(q^2,s_1,s_3)^{00-}
 +    \mF^{\mbox{\tiny Tauola}}_P(q^2, s_2,s_3  )^{00-} \, .
\eear

%
%
%

\end{document}